\documentclass[prb,twocolumn,notitlepage,superscriptaddress,longbibliography]{revtex4}
\usepackage{amsmath,amsfonts,amssymb,amsthm,epsfig,array}
\usepackage{dsfont}
\usepackage{slashed}
\usepackage{graphics}
\usepackage{float}
\usepackage{verbatim}
\usepackage{color}
\usepackage[dvipsnames]{xcolor}
\usepackage[hidelinks,colorlinks,linkcolor=blue,
citecolor=blue,urlcolor=blue]{hyperref}
\usepackage[titletoc,title]{appendix}
\usepackage{multirow}
\usepackage{bibentry}

\setcitestyle{number,sort,open={(},close={)}}

\newcommand{\Tr}{\mathop{\mathrm{Tr}}}

\newcommand{\eqnref}[1]{Eq.\,\eqref{#1}}
\newcommand{\figref}[1]{Fig.\,\ref{#1}}

\newcommand{\eq}[1]{\begin{equation} #1 \end{equation}}

\let\oldAA\AA
\renewcommand{\AA}{\text{\normalfont\oldAA}}

\newcommand{\TR}{\mathcal{T}}

\usepackage[mathscr]{euscript}
\renewcommand{\vec}[1]{{\textbf{\textit{#1}}}}

\begin{document}
\title{Momentum-Space Spin Antivortex and Spin Transport in Monolayer Pb}
\author{Kaijie Yang}
\affiliation{Department of Physics, The Pennsylvania State University, University Park, Pennsylvania 16802}
\author{Yuanxi Wang}
\affiliation{2-Dimensional Crystal Consortium, Materials Research Institute, The Pennsylvania State University, University Park, Pennsylvania 16802}
\affiliation{Department of Physics, University of North Texas, Denton, Texas 76203}
\author{Chao-Xing Liu}
\email{cxl56@psu.edu}
\affiliation{Department of Physics, the Pennsylvania State University, University Park, Pennsylvania 16802}
\begin{abstract}
	Nontrivial momentum-space spin texture of electrons can be induced by spin-orbit coupling and underpins various spin transport phenomena, such as current-induced spin polarization and spin Hall effect. In this work, we find that a nontrivial spin texture, spin antivortex, can appear at certain momenta on the $\Gamma-K$ line in 2D monolayer Pb on top of SiC. Different from spin vortex due to the band degeneracy in the Rashba model, the existence of this spin antivortex is guaranteed by the Poincar\'e-Hopf theorem and thus topologically stable. Accompanied with this spin antivortex, a Lifshitz transition of Fermi surfaces occurs at certain momenta on the $K-M$ line, and both phenomena are originated from the anticrossing between the $j=1/2$ and $j=3/2$ bands.	A rapid variation of the response coefficients for both the current-induced spin polarization and spin Hall conductivity is found when the Fermi energy is tuned around the spin antivortex. Our work demonstrates monolayer Pb as a potentially appealing platform for spintronic applications.
\end{abstract}
\maketitle

{\it Introduction $-$}
Momentum-space spin textures of electronic bands often provide an intuitive picture to understand spin transport phenomena such as current-induced spin polarization (CISP, also known as the Edelstein effect or the inverse spin galvanic effect) \cite{edelstein1990spin,kato2004current,silov2004current,sih2005spatial,zhang2015charge,li2016direct,vzutic2004spintronics} and spin Hall effect (SHE)  \cite{hirsch1999spin,sinova2004universal,murakami2003dissipationless} in spin-orbit coupled materials. The Rashba model \cite{inoue2003diffuse,trushin2007anisotropic,johansson2016theoretical,inoue2004suppression}, for example, possesses a spin texture of the vortex type, for which electron spins are oriented tangentially along the Fermi contour and form a vortex texture (also called helical texture) for each of the two spin-split bands. Under an electric field, the nonequilibrium distribution of electrons around a shifted Fermi contour carrying spin vortices leads to a net CISP. Such analytical models of spin vortices usually capture spin textures around high-symmetry $k$ points but can be insufficient in describing response functions that rely on {\it integrals} over the entire Brillouin zone in realistic materials, where complex spin textures arising far from high-symmetry $k$ points contribute dominantly. Here we introduce a new type of spin texture, a {\it spin antivortex}, with a unique movable but unremovable nature in the momentum space.
We demonstrate that spin antivortices can be hosted by atomically thin metals with strong spin orbit coupling (SOC). Such 2D metals have been realized using the confinement heteroepitaxy technique \cite{briggs2020atomically}, where metal species  intercalate underneath graphene epitaxially grown on an insulating SiC substrate. This new growth technique can produce air-stable, crystalline 2D metals at scale, with extraordinary optical \cite{steves2020unexpected} 
and transport properties \cite{briggs2020atomically} that may enable the development of new device applications of 2D materials \cite{miro2014atlas,avsar2020colloquium,das2015beyond}.
In particular, the strong SOC in 2D heavy metals, such as Pb, Pt, Sn, Au and Bi, can naturally produce spin-split bands when combined with the inversion symmetry breaking guaranteed by the graphene-metal-SiC architecture.

\begin{figure*}[t]
	\centering
	\includegraphics[width=\textwidth]{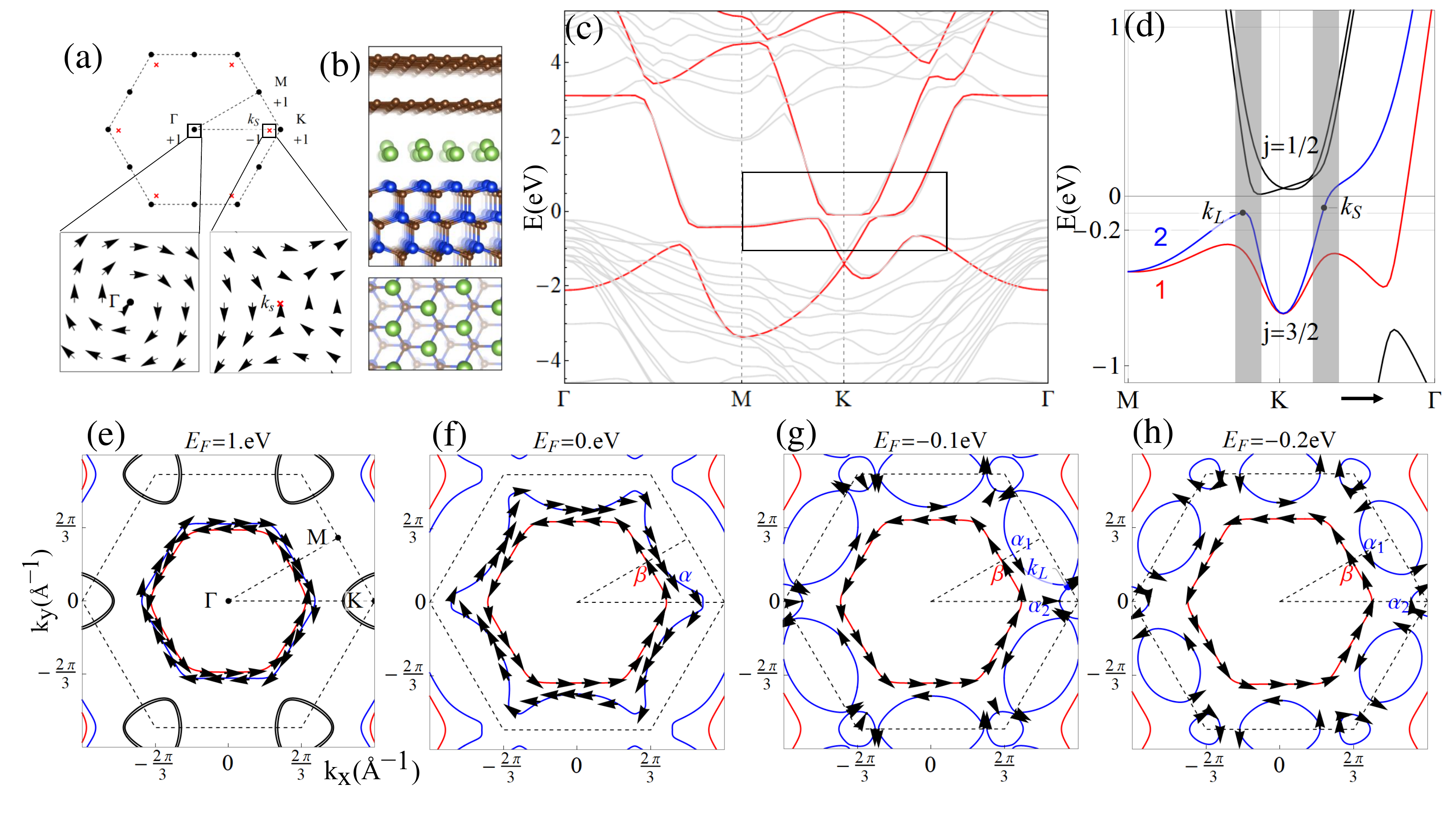}
	\caption{
		(a) The locations of spin vortices (black dots) and antivortices (red crosses) in the first BZ and the spin texture for spin vortex around the $\Gamma$ point with a winding number $+1$ and the antivortex around the $k_\text S$ point with a winding number $-1$.
		(b) The lattice structure of the intercalated monolayer Pb between SiC substrate and graphene.
		(c) The electronic band structure without SOC from first principal calculation (gray lines) and from the maximal localized Wannier function (MLWF) (red lines).
		(d) The band spectrum from MLWF with SOC for the energy range of the black box in (b).  The gray region is the anticrossing regime of $j=1/2$ and $j=3/2$ bands. $k_\text L$ and $k_\text S$ label the locations of the Lifshitz transition point and spin antivortex.
		(e)$-$(h) Fermi surfaces and spin texture of band 1 and 2 for $E_F=1.0,0.0,-0.1,-0.2$ eV.
		$\alpha$($\beta$) are hole pockets for band 1(2) around $\Gamma$ before the Lifshitz transition. Black lines are electron Fermi pockets. $\alpha_1$($\alpha_2$) are Fermi surface pockets for band 2 around $ M(K)$ after Lifshitz transition.
	}
	\label{fig:spectrum}
\end{figure*}

In this work, we study the spin texture for the 2D Pb monolayer and find that besides the spin vortices around high-symmetry momenta $\Gamma$, $M$ and $K$, spin antivortices with the opposite winding numbers exist at non-high-symmetry momenta along the $\Gamma-K$ lines, as labelled by six red crosses in \figref{fig:spectrum}(a). Unlike the Rashba model, in which spin vortices are induced by band degeneracy, the spin antivortices in this system are guaranteed by the Poincar\'e-Hopf theorem, and thus reveal a movable but locally unremovable nature. A Lifshitz transition of the Fermi surface along K$-$M accompanies the emergence of spin antivortices in a similar energy range. To quantitatively predict the spectroscopic signatures of the spin antivortices and  Lifshitz transition, we model their effect on spin transport by combining density functional theory (DFT)\cite{guo2005ab,guo2008intrinsic,freimuth2010anisotropic,lowitzer2011extrinsic,gradhand2010extrinsic,gradhand2010spin} with the Green's function formalism to show that they induce a rapid change of CISP and SHE when the Fermi energy is tuned to align with them. The effect of short-range disorder scattering is also discussed by including the vertex correction in the Green's function formalism. Our work will guide  experimental studies on spin phenomena and pave the way to the spintronic applications of 2D heavy metals.

{\it Lifshitz transition and momentum-space spin antivortex in monolayer Pb $-$}
We start from the ground state lattice structure and electronic structure of monolayer Pb on top of SiC substrate. The Pb atoms form a triangular lattice described by the $C_{3v}$ symmetry group that can be generated by a threefold rotation and an in-plane mirror. Inversion symmetry is broken due to the local environment, as shown in \figref{fig:spectrum}(b).
The electronic structure of this system at the DFT level (see Supplemental Material \cite{SM} Sec.\MakeUppercase{ \romannumeral 1}.B for details) without and with SOC are, respectively, shown in ~\figref{fig:spectrum} (c) and 1(d). Focusing on the $-1$ to $1$eV energy range near the Fermi energy, the two strongly dispersive bands are mainly characterized by Pb $p_x$ and $p_y$ orbitals; the other weakly dispersive band anticrossing with the $p_x$ and $p_y$ bands is of $p_z$ orbital character from both Pb and the topmost Si layer of the SiC substrate. Band interpolation using maximally localized Wannier functions was then performed using the above four orbitals for initial projections (details in Supplemental Material \cite{SM}). The resulting Wannier-interpolated bands in ~\figref{fig:spectrum}(c) agrees well with the original DFT ones within the manifold of the four orbitals.
After introducing atomic SOC to the tight binding Hamiltonian obtained from Wannierization, we obtain low-energy bands shown in ~\figref{fig:spectrum}(d). Labeling bands with SOC by total angular momenta $j$ at K, the $j=1/2$ bands are mainly dominated by Si $p_z$ orbitals and the $j=3/2$ bands [labeled as bands 1 and 2 in \figref{fig:spectrum}(d)] come from Pb $p_x \ p_y$ orbitals; these two bands anticross around $K$ for $E_F=-0.1$ eV.

We next show the evolution of Fermi surfaces across a Lifshitz transition in \figref{fig:spectrum}(e-h), where $E_F$ is lowered from $1.0$eV to $0.0, -0.1$ and $-0.2$ eV.
At $E_F=1.0$ eV, the spin-split hole pockets $\alpha$ (blue) and $\beta$ (red) and the spin-split electron pockets around K (black) come from Pb $p_x, p_y$ orbitals.
As the Fermi energy lowers to $E_F= 0.0 $ eV, the electron pockets shrink and disappear, while the hole pockets extend towards the Brillouin Zone (BZ) boundary. As the Fermi energy decreases further, the hole pocket $\alpha$ are split into the electron pockets $\alpha_1$ and $\alpha_2$ around the M and K points, as shown in \figref{fig:spectrum}(h). A Lifshitz transition that changes the Fermi surface topology occurs at the momentum $k_\text L$ along the K$-$M line at  $E_F=-0.1 $ eV, as shown in \figref{fig:spectrum}(g).

\begin{figure}[t]
	\centering
	\includegraphics[width=\columnwidth]{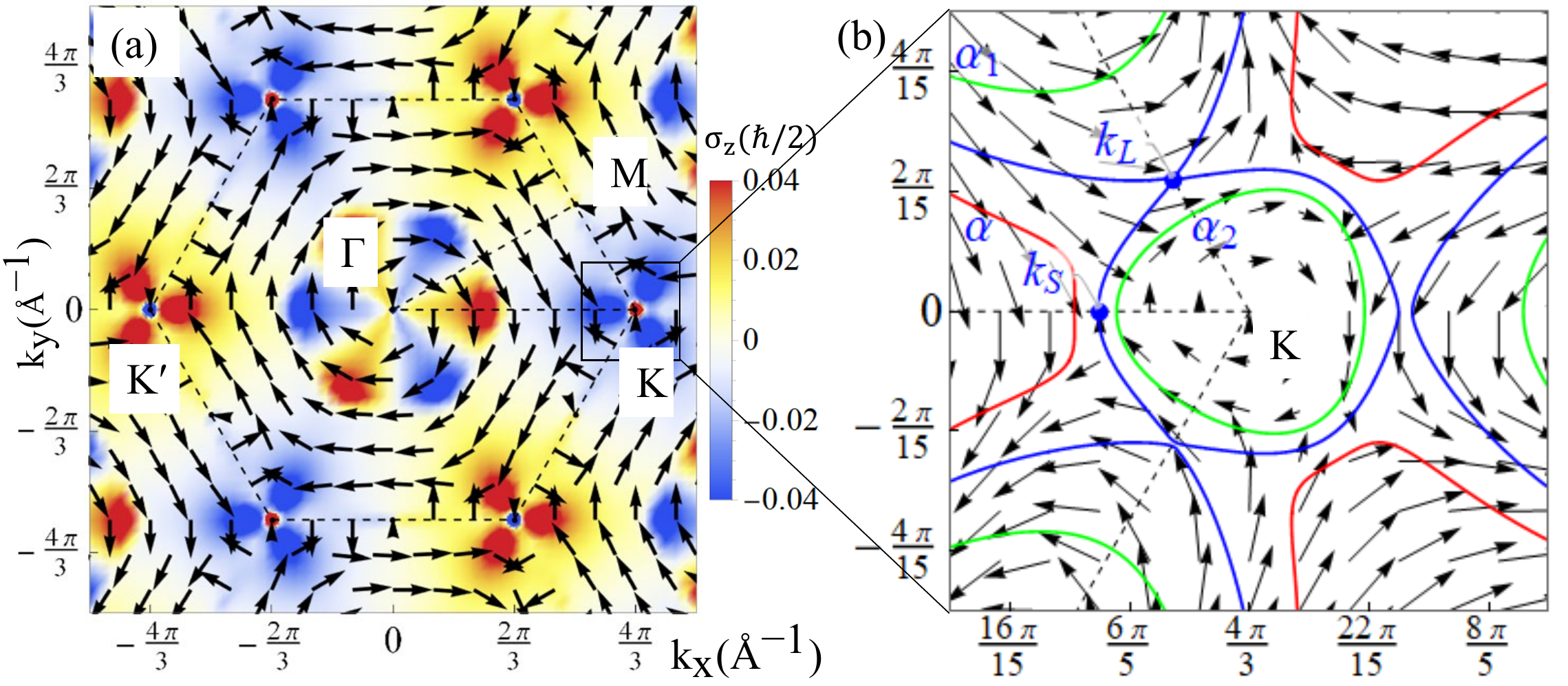}
	\caption{
		(a) Spin texture of the band 2 in the whole BZ. The arrows depict in-plane spin polarization while the background color reveals the $z$-component spin polarization. The spin vortices with winding number +1 exist at high symmetry points $\Gamma$, $M$, $K$, $K'$. 
		(b) The zoom-in of spin texture around $K$. A spin antivortex shows up on the $\Gamma-K$ line at $k_\text S$ point on Fermi surfaces around $E_F=-0.1$ eV, concurrent with the Lifshitz transition.  The red, blue, green lines are Fermi surfaces of $E_F=0.0, -0.1, -0.2$ eV respectively.
	}
	\label{fig:spintexture}
\end{figure}

Also evolving with lowering Fermi energies along \figref{fig:spectrum}(e$-$h) are spin textures of the Fermi pockets. At $E_F= 1.0 $ eV, the spin textures of two hole pockets are of Rashba-type. As the Fermi energy lowers, the spin texture of the inner  hole pocket $\beta$ (red) remains the same while dramatic changes occur for the outer hole pocket $\alpha$ (blue) that experiences the Lifshitz transition. By comparing the spin direction of the blue Fermi pocket at $E_F= 0.0 $ and $E_F= -0.2 $ eV, one notices that the spin direction of the Fermi pocket $\alpha_1$ around $M$
at $E_F= -0.2 $ eV follows that at $E_F= 0.0 $ eV, while the spin direction of the Fermi pocket $\alpha_2$ around K at $E_F= -0.2 $ eV reverses its sign, as compared to that at $E_F= 0.0 $ eV. To see this feature more clearly, we focus on the spin texture of band 2 in the whole BZ, as shown in \figref{fig:spintexture}(a), in which the arrows show the in-plane spin directions and the colors indicate the $z$-component spin. It is clear that the in-plane spin forms vortices around $\Gamma$, $M$, $K$, and $K'$. Zooming in on the spin texture of band 2 around K in \figref{fig:spintexture}(b), we find that, besides the vortex around K, another antivortex centered at $k_\text S$ appears along the $\Gamma-K$ line.
The lines $\alpha$, $\alpha_1$ and $\alpha_2$ in \figref{fig:spintexture}(b) show the Fermi surfaces at $E_F= 0.0 $ eV and $E_F= -0.2 $ eV, respectively, and we indeed see a sign change of the spin direction for the momenta at the left and right sides of $k_\text S$ along the $\Gamma-$K line between these two Fermi surfaces, consistent with the Fermi pocket plot in \figref{fig:spectrum}(f) and 1(h). 

The existence of the spin antivortex can be viewed as a consequence of the Poincar\'e-Hopf theorem of a tangential vector field on a compact manifold \cite{fulton2013algebraic}. The BZ is a torus and we only focus on the in-plane spin component, which can be regarded as a tangential field.
According to the Poincar\'e-Hopf theorem,
the total of winding numbers around spin vortex is the Euler number of the torus, namely zero. The winding numbers of spin vortices at six high symmetry momenta, namely $\Gamma$, $K$, $K'$ and three $M$ points, are all $+1$, as shown in \figref{fig:spintexture}(a). On the other hand, there are six antivortices due to the combination of $C_3$ and $\TR$ symmetries, each with winding number $-1$, as shown in \figref{fig:spectrum}(a).
Thus, the total winding number in the whole BZ vanishes, as required by the Poincar\'e-Hopf theorem. This analysis suggests that, unlike a spin vortex around high symmetry momenta, each spin antivortex is movable along the $\Gamma-K$ lines,  but is locally stable due to its topological nature. The difference between spin antivortex and other spin texture, such as spin vortex in the Rashba model, is further discussed in Sec.\MakeUppercase{\romannumeral 2}.C of Supplemental Material\cite{SM} . In addition, it is shown that the location of spin antivortex can be controlled by an external gate voltage, as discussed in Sec.\MakeUppercase{\romannumeral 2}.B of Supplemental Material\cite{SM}.

It is interesting to notice that both the Lifshitz transition and the spin antivortex occur approximately around $E_F \approx -0.1 $ eV, as shown in \figref{fig:spintexture}(b). This is because both phenomena are related to the anticrossings between the $j=1/2$ bands from $p_z$ orbitals and the $j=3/2$ bands from $p_{x,y}$ orbitals around $K$ or $K'$ , as shown by the gray regime in \figref{fig:spectrum}(d).
We provide more detailed analysis on how the band anticrossing can induce the Lifshitz transition and spin antivortex in Supplemental Material\cite{SM} Sec.\MakeUppercase{\romannumeral 2}.B.

\begin{figure*}[t]
	\centering
	\includegraphics[width=\textwidth]{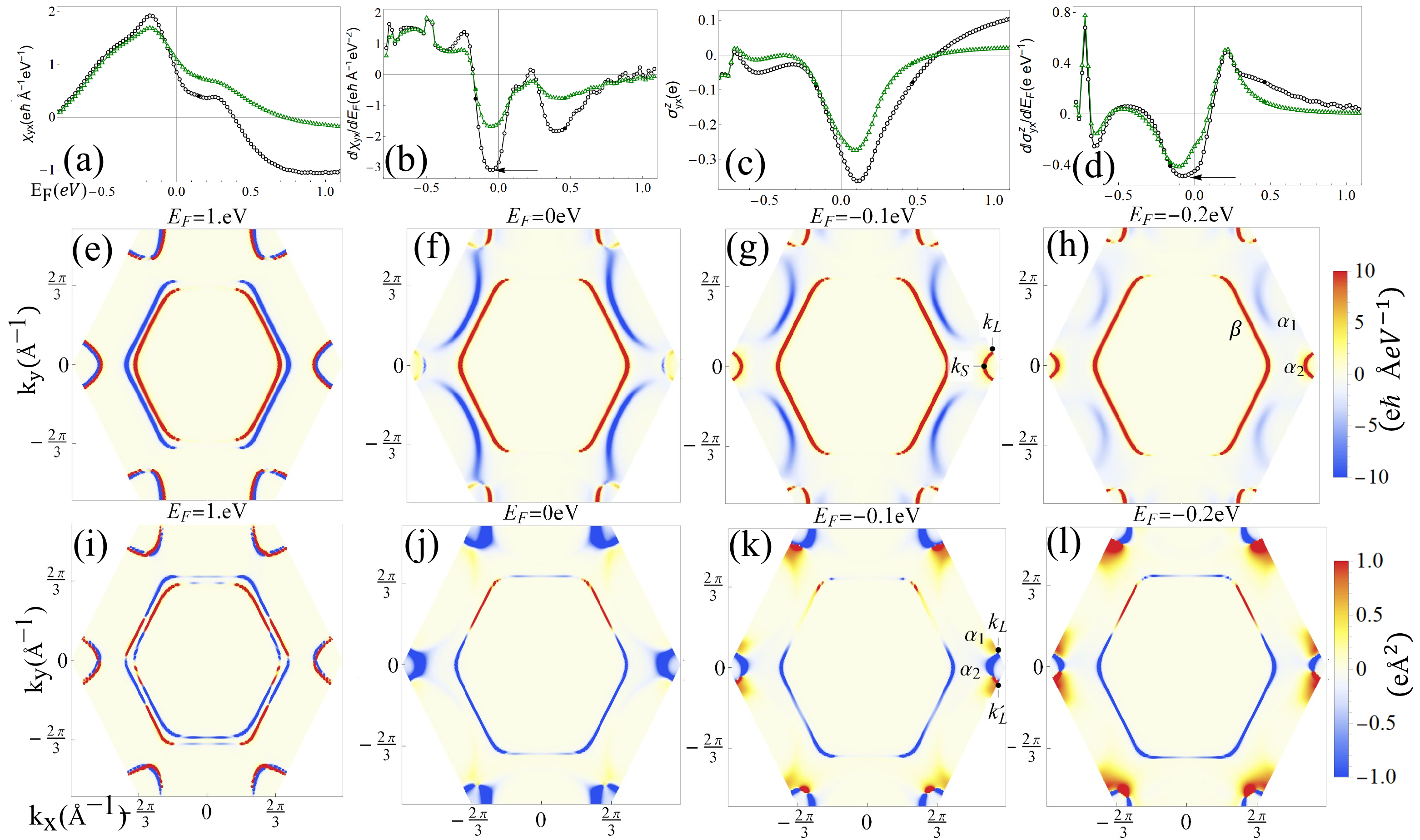}
	\caption{
		(a), (b), (c) and (d) show $\chi_{yx}$, $d \chi_{yx} / d E_F$, $\sigma^z_{yx}$, and $d\sigma^z_{yx}/dE_F$, respectively, as a function of $E_\text F$, in the clean case (black lines with circles) and with disorder scattering (green lines with triangles) for different $E_F$.
		(e)$-$(h) and (i)$-$(l) show the momentum resolved contribution $\chi_{ij}(\vec k, E_\text F)$ to the CISP and $\sigma^k_{ij}(\vec k,E_\text F)$ to the SHE, respectively, over the BZ for $E_F=1.0,0.0,-0.1,-0.2$ eV (without vertex correction).
	}
	\label{fig:cisp}
\end{figure*}

{\it Current-induced spin polarization and spin Hall effect $-$}
The Lifshitz transition and spin antivortex can be experimentally verified by their spectroscopic signatures: they can, in principle, be extracted through the spin-resolved angular-resolved photoemission spectroscopy \cite{sobota2021angle}. Here we focus on the spin transport phenomena of CISP and SHE, which are described by the response equations, $S_k = \sum_i \chi_{ki} E_i$ for CISP and $J^k_j = \sum_i \sigma^k_{ji} E_i$ for SHE, respectively. Here $S_k$ is the spin operator with $k=x,y,z$, $v_j$ is the velocity operator with $i,j=x,y$ and $J^k_j = \{ S_k, v_j\} / 2$ is the spin current operator.
Based on the threefold rotation and in-plane mirror symmetries
, the nonzero in-plane current response coefficients for $x$-direction electrical field are $\chi_{yx}$ for CISP and $\sigma^z_{yx}, \sigma^x_{xx}=-\sigma^y_{yx}$ for SHE from the Neumann's principles \cite{neumannprinciple}.

The detailed form of the response coefficients $\chi_{ij}$ and $\sigma^k_{ji}$ can be derived from the standard linear response theory and are given by
\eq{
	\begin{split}
		& \chi_{ij}(E_\text F) = \int \frac{d^2 \vec k}{(2\pi)^2} \chi_{ij}(\vec k, E_\text F) \\
		& = -\frac{1}{2\pi} \int \frac{d^2 \vec k}{(2\pi)^2} \Tr S_i G^R(\vec k, E_F) \Gamma_j(\vec k) G^A (\vec k, E_F)
	\end{split}
	\label{eq:cisp}
}
and
\eq{
	\begin{split}
		& \sigma^k_{ij}(E_\text F) = \int \frac{d^2 \vec k}{(2\pi)^2} \sigma^k_{ij}(\vec k, E_\text F) \\
		& = -\frac{1}{2\pi} \int \frac{d^2 \vec k}{(2\pi)^2} \Tr J^k_i(\vec k) G^R(\vec k, E_F) \Gamma_j(\vec k) G^A (\vec k, E_F)
	\end{split}
	\label{eq:she}
	\raisetag{35pt}
}
where $G^R$ and $G^A$ are the retarded and advanced Green's functions, $\Tr$ is the trace taken over band indices, and the vertex operator $\Gamma_j(\vec k)$ is defined as $\Gamma_j (\vec k, E) = J_j(\vec k) + \gamma_j(E)$ with the current operator $J_j = - e v_j$ and
\eq{
	\gamma_j(E) = n_i V_0^2 \int \frac{d^2 \vec l}{(2\pi)^2} G^R(\vec l, E)( J_j (\vec l, E)  + \gamma_j (E) )G^A(\vec l, E).
}
Here, $\gamma_j$ describes the vertex correction to the current operator from the ladder diagram of disorder scattering with the disorder density $n_i$ and scattering strength $V_0$ \cite{mahan2013many}.
We apply the linear response formalism to the above-mentioned tight-binding model parametrized using Wannier functions;  details of the numerical technique are discussed in Supplemental Material\cite{SM} Sec.\MakeUppercase{ \romannumeral 1 }.

\figref{fig:cisp} summarizes our main numerical results for $\chi_{yx} (E_\text F)$ for CISP and $\sigma^z_{yx} (E_\text F)$ for SHE, respectively. Here the green line is for the case without the vertex correction, namely $\Gamma_j (\vec k, E) = J_j(\vec k)$, and the black line with vertex correction. We first analyze $\chi_{yx}(E_\text F)$ and plot its momentum resolved contribution $\chi_{yx}(\vec k,E_\text F)$ of \eqnref{eq:cisp} in \figref{fig:cisp}(e)$-$3(i). We find that the main contribution to CISP comes from the Fermi surface since $G^R(\vec k, E_F)G^A(\vec k, E_F)$ in \eqref{eq:cisp} can be approximated by $\delta(E_F - \epsilon_{nk})$ in the dilute disorder limit. At $E_F= 1.0 $ eV, due to the opposite spin texture for the two hole pockets, they contribute oppositely to CISP, as shown in \figref{fig:cisp}(e). The outer Fermi pocket will be dominant and thus gives rise to the negative sign of $\gamma_{yx}$, resembling the CISP contribution in the standard Rashba model \cite{inoue2003diffuse}. Near the Lifshitz transition, a dramatic change occurs to the outer Fermi pocket $\alpha_2$ around $K$ and $K'$. When lowering the Fermi energy from $E_F=0.0$ to $-1.0$ eV, one can clearly see that the Fermi pocket $\alpha_2$ around  $K$ and $K'$ varies from blue (negative) to red (positive), as shown in \figref{fig:cisp}(f) and 3(g).
Physically, this sign change directly reflects the presence of spin antivortices, as the spin direction on the Fermi pocket $\alpha_2$ around  $K$ and $K'$ changes its sign across the spin antivortex. In addition, the contribution to CISP from the Fermi pocket $\alpha_1$ around M also decreases due to the reduction of relaxation time (See Sec.\MakeUppercase{ \romannumeral 2 }.C of Supplemental Material\cite{SM}). Combining the above factors leads to (i) the positive sign of $\chi_{yx}$ in this Fermi energy range ($E_F<  0.3$ eV) and (ii) a rapid increase of $\chi_{yx}$ around $E_F\approx -0.1 $ eV as $E_F$ decreases. To see this rapid change more clearly, we take a derivative of $\chi_{yx}$ with respect to $E_F$, as shown in \figref{fig:cisp}(b). The peak in the amplitude of $d\chi_{yx}/dE_F$ indeed occurs around $E_F\approx -0.1 $ eV, as shown by the arrow in \figref{fig:cisp}(b), and thus is induced by the Fermi surface crossing the antivortex.
The SHE also reveals a rapid change around $E_F \approx -0.1 $ eV, as shown by a similar peak appears for $d\sigma^z_{yx} / d E_F$ in \figref{fig:cisp}(d).
This is because the contribution to SHE around the Fermi pockets $\alpha_1$ close to the Lifshitz transition point $k_\text L$ and that around $\alpha_2$ close to $k_\text L'$ abruptly changes sign from negative (blue) to positive (red) when lowering the Fermi energy from $E_F=0.0$ to $-0.1$ eV, as shown in \figref{fig:cisp}(j) and 3(k). Detailed analysis suggests that the sign change of contribution around $k_\text L$ and $k_\text L^\prime$ is more closely related to the Lifshitz transition due to the fact that the SHE mainly is contributed from interband matrix elements rather than the intraband Fermi surface contribution, as discussed in Sec.\MakeUppercase{ \romannumeral 2 }.D of Supplemental Material\cite{SM}. The influence of disorder scattering is also evaluated through the vertex correction, as shown by the dashed lines in \figref{fig:cisp}(a), 3(b) and 3(c), 3(d). We generally find both the values of $\chi_{yx}$ and $\sigma^z_{yx}$, as well as $d \chi_{yx} / d E_F$ and $d\sigma^z_{yx}/dE_F$, around $E_F=1.0$ eV are enhanced by disorder scattering, as discussed in details in Sec.\MakeUppercase{\romannumeral 2}.C and \MakeUppercase{\romannumeral 2}.D of Supplemental Material\cite{SM}.

{\it Conclusion and discussion $-$}
In conclusion, the spin antivortices and Lifshitz transition induced by the anticrossing between the $p_z$ and $p_{x,y}$ bands strongly affect spin transport phenomena in the monolayer Pb on SiC.  By choosing appropriate parameters (see Sec.\MakeUppercase{\romannumeral 2}.E of Supplemental Material\cite{SM}), we find $\sigma^z_{yx}$ is of the order $10^{-1} e$ in our model and the corresponding spin Hall angle is $\sim 10^{-2}$, comparable to the existing experimentally measured values \cite{morota2011indication,mosendz2010detection,ando2010inverse}. Based on the same values, we find the variation of $\sigma^z_{yx}$ across the band anticrossing regime is around $\sim 0.3e$, and thus should be measurable by tuning the carrier density of 2D Pb in experiments.
For CISP, $\chi_{yx}/\hbar$ is of the order $10^{-8} \ \text{nm}^{-2} \text V^{-1} \text m$ and its variation is $\sim 2 \times 10^{-8} \ \text{nm}^{-2} \text V^{-1} \text m$ in the band anticrossing regime. The charge-to-spin conversion efficiency $2 e v_f \chi_{yx} / \hbar \sigma_{xx}$ is $\sim 0.1$, where $v_f$ is the Fermi velocity and $\sigma_{xx}$ is the longitudinal conductivity. This efficiency is close to that proposed and measured of graphene on a transition-metal dichalcogenide \cite{ghiasi2019charge,offidani2017optimal}.
While Pb films have been grown on top of SiC with different growth methods
\cite{yurtsever2016effects,chen2020growth,wang2021manipulation,hupalo2011metal,liu2015selective}, our theory suggests the monolayer Pb as an excellent platform for the study of spin transport phenomena and spintronic applications.

{\it Acknowledgments--.}
We thank V. Crespi, J. Robinson, N. Samarth, W. Yanez and J. Zhu for helpful discussions. This work is mainly supported by the Penn State MRSECCenter for Nanoscale Science via NSF Grant No. DMR-2011839. C.-X.~L. also acknowledges the support of the Office of Naval Research (Grant No. N00014-18-1-2793).



\end{document}


\title{Supplementary Materials of "Momentum-space Spin Anti-vortex and Spin Transport in Monolayer Pb"}
\author{Kaijie Yang}
\affiliation{Department of Physics, the Pennsylvania State University, University Park, PA 16802}
\author{Yuanxi Wang}
\affiliation{2-Dimensional Crystal Consortium, Materials Research Institute, The Pennsylvania State University, University Park, PA 16802}
\affiliation{Department of Physics, University of North Texas, Denton, TX 76203}
\author{Chao-Xing Liu}
\email{cxl56@psu.edu}
\affiliation{Department of Physics, the Pennsylvania State University, University Park, PA 16802}

\maketitle

\section{Numerical linear response of spin transport with short range disorder}

\subsection{formalism}
\label{sec:formalism}

The formalism for CISP $\chi_{ij}$ and SHE $\sigma^k_{ij}$ in orbital basis are derived from the linear response theory \cite{mahan2013many} and are given in the main text by
\begin{equation}
	\chi_{ij} = -\frac{1}{2\pi} \int \frac{d^2 \vec k}{(2\pi)^2} \Tr S_i G^R(\vec k, E_f) \Gamma_j(\vec k,E_f) G^A (\vec k, E_f)
	\label{eq:cisp}
\end{equation}
and
\begin{equation}
	\sigma^k_{ij} = -\frac{1}{2\pi} \int \frac{d^2 \vec k}{(2\pi)^2} \Tr J^k_i(\vec k) G^R(\vec k, E_f) \Gamma_j(\vec k,E_f) G^A (\vec k, E_f)
	\label{eq:she}
\end{equation}
where the Green's functions are defined as
\begin{equation}
	G^R(\vec k, E) = (E-H(\vec k)-\Sigma^R(E))^{-1}.
	\label{eq:green}
\end{equation}
The integrand
\begin{equation}
	\chi_{ij}(\vec k) = -\frac{1}{2\pi} \Tr S_i G^R(\vec k, E_f) \Gamma_j(\vec k,E_f) G^A (\vec k, E_f)
	\label{eq:cispcontribution}
\end{equation}
is defined as the contribution over the Brillouin zone (BZ) to CISP.  The contribution over BZ to SHE is also defined similarly for the integrand \begin{equation}
	\sigma^k_{ij} (\vec k) = -\frac{1}{2\pi} \Tr J^k_i(\vec k) G^R(\vec k, E_f) \Gamma_j(\vec k,E_f) G^A (\vec k, E_f).
	\label{eq:shecontribution}
\end{equation}

The self energies $\Sigma^R(\vec k, E)$ in the Green's functions come from short range disorder.  The disorder has the $\delta$-like potential $V(r) = \sum_i V_0 \delta(\vec r - \vec R_i)$ and the density $n_i$ where $V_0$ is the disorder strength and $R_i$ is the disorder position.  After disorder average, the self energies in the self-consistent Born approximation are
\begin{equation}
	\Sigma^R(E) = n_i V_0^2 \int \frac{d^2 \vec k'}{(2\pi)^2} G^R(\vec k', E).
	\label{eq:selfenergy}
\end{equation}

The vertex correction $\Gamma_j(\vec k, E) = J_j(\vec k) + \gamma_j(E)$ is written in the ladder approximation as in the main text
\begin{equation}
	\gamma_j(E) = n_i V_0^2 \int \frac{d^2 \vec l}{(2\pi)^2} G^R(\vec l, E)( J_j (\vec l)  + \gamma_j (E) )G^A(\vec l, E).
	\label{eq:vertexcorrection}
\end{equation}
The $\gamma_j(E)$ is independent of momenta because the $\delta$-like disorder potential $V({\bf r})$ becomes a constant $V(\vec k- \vec l)$ for all transferred momenta after the transform to the momentum space.

When the disorder is dilute, the formalism transformed to energy eigenspace of Hamiltonian can be reduced to the Fermi surfaces. The self-consistent Born approximation for self energies is reduced to first-order Born approximation and reads
\begin{equation}
	\Im \Sigma^R(\vec k n, E) = n_i V_0^2 \sum_{n'}  \int \frac{d^2 \vec k'}{(2\pi)^2} (-\pi) \delta(E-\epsilon_{\vec k' n'}) \vert \langle \vec k n \vert \vec k' n'  \rangle \vert^2.
	\label{eq:selfenergyfs}
\end{equation}
The $\ket{\vec k n}$ are the eigen states of Hamiltonian and $\epsilon_{\vec k n}$ are the corresponding eigen energies.
The imaginary part of the bare Green's function in energy eigenspace $G^R_0(\vec k n, E) = (E - \epsilon_{\vec k n} +i 0^+)^{-1}$ is approximated by $-\pi \delta(E-\epsilon_{\vec k n})$.  The $\delta$-function shows the contribution to self-energies come from the Fermi surface.  The off-diagonal parts of self energies in eigen-energy space are dropped.
Under the time reversal symmetry $\TR$, the diagonal self-energies become
\begin{equation}
\begin{split}
		\Im \Sigma^R(\vec k n, E) &= n_i V_0^2 \sum_{n'}  \int \frac{d^2 \vec k'}{(2\pi)^2} (-\pi) \delta(E-\epsilon_{\vec k' n'}) \vert \langle \vec k n \vert \TR^\dagger \TR\vert \vec k' n'  \rangle \vert^2.\\
		& =  n_i V_0^2 \sum_{n'}  \int \frac{d^2 \vec k'}{(2\pi)^2} (-\pi) \delta(E-\epsilon_{\vec k' n'}) \vert \langle -\vec k' n \vert - \vec k n'  \rangle \vert^2 \\
		& = \Im \Sigma^R( - \vec k n, E).
\end{split}
\end{equation}
The relaxation time is
\begin{equation}
	\tau_{\vec k n}(E) = \frac{-1}{ 2\Im \Sigma^R(\vec k n, E)}.
\end{equation}

To evaluate the CISP in Eq. (\ref{eq:cisp}), we consider the approximation
\begin{equation}
	G^R(\vec k n, E) G^A(\vec k n', E) \approx \delta_{nn'} \delta(E-\epsilon_{\vec k n}) \frac{\pi}{-\Im \Sigma^R(\vec k n, E)},
	\label{eq:fsapproximation}
\end{equation}
in which the off-diagonal part $n \neq n'$ is of the order $ \Im \Sigma^R / (\epsilon_{\vec k n} - \epsilon_{\vec k n'})$, as compared to diagonal part $n = n'$, and thus is neglected. Under this approximation, the response coefficient $\chi_{ij}$ of the CISP in energy eigenspace reads
\begin{equation}
	\chi_{ij} =  -\frac{1}{ 2\pi } \sum_n \int \frac{d^2 \vec k}{(2\pi)^2} \langle \vec k n \vert S_i \vert \vec k n \rangle \langle \vec k n \vert \Gamma_j(\vec k, E_f) \vert \vec k n \rangle \delta(E_f -\epsilon_{\vec k n}) \frac{\pi}{-\Im \Sigma^R(\vec k n, E_f)}.
	\label{eq:cispfs}
\end{equation}
Below we will also consider the integrand
\begin{equation}
	\chi_{ij}(\vec k) =  -\frac{1}{ 2\pi }  \langle \vec k n \vert S_i \vert \vec k n \rangle \langle \vec k n \vert \Gamma_j(\vec k, E_f) \vert \vec k n \rangle \delta(E_f -\epsilon_{\vec k n}) \frac{\pi}{-\Im \Sigma^R(\vec k n, E_f)},
	\label{eq:cispfscontribution}
\end{equation}
which reveals the momentum-resolved contribution around the Fermi surfaces of the band n.
Under the similar approximation, the vertex correction reads
\begin{equation}
	\gamma_j(\vec k n, E) = n_i V_0^2 \sum_m \int \frac{d^2 \vec l}{(2\pi)^2} ( J_j(\vec lm) + \gamma_j(\vec l m,E)) \abs{\bra{\vec k n} \vec l m \rangle}^2  \delta(E-\epsilon_{\vec l m}) \frac{\pi}{-\Im \Sigma^R(\vec l m, E)}
	\label{eq:vertexcorrectionfs}
\end{equation}
where $\gamma_j(\vec k n, E)=\bra{\vec k n}\gamma_j(E) \ket{\vec k n}$ and $J_j(\vec k n)=\bra{\vec k n}J_j(\vec k) \ket{\vec k n}$.  After obtaining $\gamma_j(\vec k n,E)$, the off-diagonal part of the vertex correction is
\begin{equation}
	\bra{\vec k n}\gamma_j(E) \ket{\vec k n'} = n_i V_0^2 \sum_m \int \frac{d^2 \vec l}{(2\pi)^2} ( J_j(\vec lm) + \gamma_j(\vec l m,E)) \bra{\vec k n} \vec l m \rangle \langle lm \ket{\vec k n'} \delta(E-\epsilon_{\vec l m}) \frac{\pi}{-\Im \Sigma^R(\vec l m, E)}.
\end{equation}
The vertex correction can also be calculated for the spin operator in CISP as
\begin{equation}
	\bra{\vec k n}\gamma^S_i(E) \ket{\vec k n} = n_i V_0^2 \sum_m \int \frac{d^2 \vec l}{(2\pi)^2} ( \bra{\vec l m} S_i \ket{\vec l m} + \bra{\vec l m}\gamma^S_i(E) \ket{\vec l m}) \abs{\bra{\vec k n} \vec l m \rangle}^2 \delta(E-\epsilon_{\vec l m}) \frac{\pi}{-\Im \Sigma^R(\vec l m, E)}.
\end{equation}
The spin operator with the vertex correction is
\begin{equation}
 \bra{\vec k n}\tilde S_i (E)\ket{\vec k n} = \bra{\vec k n} S_i + \gamma^S_i(E) \ket{\vec k n}.
 \label{eq:spinvertexcorrectionfs}
\end{equation}

For SHE,  \eqnref{eq:she} is divided into diagonal parts and off diagonal parts in energy eigenspace as
\begin{equation}
	\sigma^k_{ij} = \sigma^D + \sigma^{OD}.
\end{equation}
The diagonal (intraband) contribution is
\begin{equation}
	\sigma^D = -\frac{1}{ 2\pi } \sum_n \int \frac{d^2 \vec k}{(2\pi)^2} \langle \vec k n \vert J^k_i \vert \vec k n \rangle \langle \vec k n \vert \Gamma_j \vert \vec k n \rangle \delta(E_f -\epsilon_{\vec k n}) \frac{\pi}{-\Im \Sigma^R(\vec k n, E_f)}.
\end{equation}
Under $\TR$, the diagonal part becomes
\begin{equation}
\begin{split}
		\sigma^D & = -\frac{1}{ 2\pi } \sum_n \int \frac{d^2 \vec k}{(2\pi)^2} \langle - \vec k n \vert  (\TR J^k_i \TR^\dagger) \vert - \vec k n \rangle^* \langle - \vec k n \vert (\TR \Gamma_j \TR^\dagger) \vert -\vec k n \rangle^* \delta(E_f -\epsilon_{\vec k n}) \frac{\pi}{-\Im \Sigma^R(\vec k n, E_f)} \\
		& = -\frac{1}{ 2\pi } \sum_n \int \frac{d^2 \vec k}{(2\pi)^2} \langle \vec k n \vert  ( J^k_i ) \vert  \vec k n \rangle \langle \vec k n \vert (- \Gamma_j ) \vert \vec k n \rangle \delta(E_f -\epsilon_{-\vec k n}) \frac{\pi}{-\Im \Sigma^R(-\vec k n, E_f)} \\
		& = -\sigma^D.
\end{split}
	\label{eq:shetimereversal}
\end{equation}
So the intraband contribution to SHE is zero.  The off diagonal part of the real SHE conductivity reads
\begin{equation}
	\sigma^k_{ij} = \sigma^{OD} = -\frac{1}{2\pi} \sum_{n \neq n'} \int \frac{d^2 \vec k}{(2\pi)^2} (-) \Im(  G^R(\vec k n, E_f) G^A(\vec k n',E_f))\Im (\langle \vec k n' \vert J^k_i \vert \vec k n \rangle \langle \vec k n \vert \Gamma_j(\vec k,E_f) \vert \vec k n' \rangle)
\end{equation}
where the term $\Re( G^R(\vec k n, E) G^A(\vec k n',E))$ is neglected because it is of the order $\Im( \Sigma^R)^2 / (\epsilon_{\vec k n}-\epsilon_{\vec k n'})^2$ compared to the $\Im (  G^R(\vec k n, E) G^A(\vec k n',E))$.  Adopting the approximation
\begin{equation}
	\Im ( G^R(\vec k n, E) G^A(\vec k n',E)) \approx (-\pi)(\delta(E - \epsilon_{\vec k n}) + \delta(E- \epsilon_{\vec k n'})) \frac{1}{\epsilon_{\vec k n} - \epsilon_{\vec k n'}},
	\label{eq:shefsapproximation}
\end{equation}
we find the SHE from  the Fermi surface contribution is
\begin{equation}
	\sigma^k_{ij} = -\frac{1}{2\pi} \sum_{n \neq n'} \int \frac{d^2 \vec k}{(2\pi)^2} 2 \pi \delta(E_f - \epsilon_{\vec k n}) \frac{1}{\epsilon_{\vec k n} - \epsilon_{\vec k n'}} \Im (\langle \vec k n' \vert J^k_i \vert \vec k n \rangle \langle \vec k n \vert \Gamma_j(\vec k,E_f) \vert \vec k n' \rangle).
	\label{eq:shefs}
\end{equation}
We also define the momentum-resolved contribution to the SHE as
\begin{equation}
	\sigma^k_{ij} (\vec k n) = -\frac{1}{2\pi} \sum_{n'} 2 \pi \delta(E_f - \epsilon_{\vec k n}) \frac{1}{\epsilon_{\vec k n} - \epsilon_{\vec k n'}} \Im (\langle \vec k n' \vert J^k_i \vert \vec k n \rangle \langle \vec k n \vert \Gamma_j(\vec k,E_f) \vert \vec k n' \rangle)
	\label{eq:shefscontribution}
\end{equation}
at the Fermi surface of the band n.

In our numerical simulations, we directly evaluate \eqnref{eq:cisp} and \eqnref{eq:she} over the first Brillouin zone. The numerical results can be well understood from the simplified expressions of \eqnref{eq:cispfs} and \eqnref{eq:shefs}, which only depends on the contribution from the Fermi energy.


\subsection{Constructing the Tight Binding Model from DFT}

Density functional theory (DFT) calculations were based on the relaxed structure of monolayer Pb on SiC, where Pb coverage is 100\%, i.e. one Pb per surface Si atom. We focus on this surface phase here since it is the thermodynamic ground state in a wide range of Pb chemical potentials compared with competing surface phases where Pb coverage is partial. Within candidate structures satisfying 100\% coverage, the registry where each Pb is directly on top of a surface Si site is preferred against the two other high-symmetry registries (Pb projecting onto the topmost C site or a ``hollow'' site, as defined in \refcite{briggs2020atomically}). The relaxed Pb structure features one out of every three Pb atom buckling out-of-plane. $C_{3v}$ symmetry is preserved even when this buckling is considered, since our supercell was chosen to be minimal and includes just three Pb atoms. 

For DFT-level structural relaxations, the Perdew-Burke-Ernzerhof (PBE) parametrization of the exchange-correlation functional at the generalized gradient approximation level was used \cite{Perdew1996, Perdew1997}. We employed projector augmented wave pseudopotentials \cite{Blochl1994, Kresse1999b} as implemented in the Vienna Ab-initio Simulation Package (VASP) \cite{Kresse1996}. All relaxations were performed with the semi-empirical DFT-D3 van der Waals correction scheme \cite{Grimme2010} and with a force threshold of 0.01~eV/\AA . Brillouin zone sampling densities were set to be equivalent to $12\times12\times1$ for a minimal SiC surface unit cell with the in-plane lattice constant $a=3.19$\AA. The SiC slab thickness was 6 SiC units, with hydrogen capping included for the other surface (opposite to the Pb-covered surface).

To construct the tight-binding model used in the main text, we adopted a structural approximation -- a relaxed minimal SiC surface unit cell with one Pb atom -- to perform preparatory self-consistency DFT calculations for Wannier interpolation. These calculations were performed using the Quantum Espresso package \cite{Giannozzi2009} with projector augmented wave pseudopotentials and with k-point grids of $12\times12\times1$, all other convergence parameters being the same as those used in the VASP calculations. Maximally localized Wannier functions and tight-binding parameters \cite{Marzari1997} were constructed using the Wannier90 code \cite{Mostofi2008}. Band disentanglement followed the standard procedure outlined in \refcite{Souza2001}: a disentanglement window was chosen to cover the entire energy where bands with all four types of orbital characters (see main text) were present; a frozen window (where Bloch state are forced to be preserved) were chosen to cover a smaller energy range where bands with \textit{exclusively} all four types of orbital characters are present.

\subsection{Numerical Realization}

This section discusses the numerical methods based on \refcite{wei2021non} to evaluate the formula in Sec. \ref{sec:formalism}.

For the units used in numerical calculation, the length unit is $k_0^{-1} = 1 \AA$ and the energy unit is $\epsilon_0=1eV$.  The units for other physical quantities are determined from these two units accordingly.  For example, the unit for CISP is $\chi_0 = k_0 e \hbar /\epsilon_0$.  The unit for $n_i V_0^2$ in \eqnref{eq:selfenergy} and \eqnref{eq:vertexcorrection} is $k_0^2 \epsilon_0^2$.  With this unit, the $n_i V_0^2$ is taken to be $0.05$ in the main text.


The first BZ integration in \eqnref{eq:cisp}, \eqnref{eq:she}, \eqnref{eq:selfenergy} and \eqnref{eq:vertexcorrection} is replaced by summation over a mesh.  The first BZ is a hexagon with three $K$ points located at
\begin{equation}
	K =  (4\pi/3 ,0) \quad K_1 =  (  4\pi/3 \times \cos 2\pi/3, 4\pi/3 \times \sin 2\pi /3) \quad K_3 =  ( 4\pi/3 \times \cos 4\pi/3,  4\pi/3 \times \sin 4\pi /3)
\end{equation}
The mesh points are taken to be $k_1 K + k_2 K_1$, $k_1 K_1 + k_2 K_2$ and $k_1 K_2 + k_2 K$, respecting $C_3$ symmetry to ensure the current conserving (See Sec. \ref{sec:currentconservation}).  The $k_1$ and $k_2$ are two real numbers in the range $[0, 1]$ with the mesh step $\Delta k = 0.005 \times 3 /4 \pi$. The corresponding energy change of adjacent mesh points in spectrum $\Delta \epsilon \sim 0.005$ is roughly an order smaller than the self-energies $\sim n_iV_0^2$.  Thus, the BZ integration  in \eqnref{eq:cisp}, \eqnref{eq:she}, \eqnref{eq:selfenergy} and \eqnref{eq:vertexcorrection} could be replaced by
\begin{equation}
	\int \frac{d^2 \vec k}{(2\pi)^2} f(\vec k) \rightarrow \sum_{\vec k_i} \frac{(\Delta k)^2 \sin 2\pi /3 }{(2\pi)^2} f(\vec k_i)
\end{equation}
where $sin(2\pi/3)$ comes from the Jacobin of a parallelogram mesh with the angle $2\pi /3$.
In total, there are $n_k \sim 10^6$ points for the mesh.

The self-consistent self-energies and Green's functions in \eqnref{eq:green} and \eqnref{eq:selfenergy} are calculated in an iterative manner. The initial values for self-energies is taken to be $- i n_i V_0^2 / 2$, same as self-energies for 2DEG.  After 12 iteration, the difference in self-energies is $\sim 10^{-5} \ll \Delta \epsilon$.  Thus, the self-energies are considered as converged to the self-consistent values.

The vertex corrections in \eqnref{eq:vertexcorrection} are solved in the following way. We move $\gamma(E)$ out of the matrix multiplication $G^R \gamma G^A$ by rewriting \eqnref{eq:vertexcorrection} in the form of the tensor product as
\begin{equation}
	\gamma_{j,mn}(E) =\sum_{m',n'} n_i V_0^2 \int \frac{d^2 \vec l}{(2\pi)^2} (G^R(\vec l, E) \otimes G^A(\vec l,E)^T)_{mn,m'n'} (J_{j,m'n'}(\vec l) +  \gamma_{j,m'n'}(E))
\end{equation}
where $n,m$ are the basis vector indices, $(G^R \otimes G^{A \ T})_{mn,m'n'} = G^R_{m,m'}G^A_{n',n}$ and $J_{j,mn}(\vec l)$ is treated as a vector with the $(m-1)\times n_b + n$th element as $\bra{\vec l m} J_j \ket{\vec l n}$. Here $n_b$ is the number of basis.  With matrix inversion, we find that the vertex correction is solved as
\begin{equation}
	\gamma_j(E) = (1 - n_i V_0^2 \int \frac{d^2 \vec k}{(2\pi)^2} (G^R(\vec k, E) \otimes G^A(\vec k,E)^T))^{-1} n_i V_0^2 \int \frac{d^2 \vec l}{(2\pi)^2} (G^R(\vec l, E) \otimes G^A(\vec l,E)^T) J_{j}(\vec l).
	\label{eq:gammatensor}
\end{equation}
The dimension of the matrices are of order $\sim O(n_k n_b^2)$. 
The basis for calculation are the Bloch states transformed from wannier orbitals used in DFT calculation. In these basis, $\gamma_j(E)$ is independent of momenta for short range disorder and only has $n_b^2$ elements compared to $n_k n_b^2$ elements in energy eigenbasis.

Then we are going to show how to deal with the the integration with the delta function $\delta(E-\epsilon_{\vec k n})$ in \eqnref{eq:cispfs}, \eqnref{eq:shefs}, \eqnref{eq:selfenergyfs} and \eqnref{eq:vertexcorrectionfs}. The delta function $\delta(E-\epsilon_{\vec k n})$ is first integrated out by radial integration as
\begin{equation}
	\int \frac{d^2 \vec k}{(2\pi)^2} \delta(E-\epsilon_{\vec k n}) f(\vec k) \rightarrow \int \frac{k(\theta) d\theta}{(2\pi)^2} \frac{1}{\vert \partial\epsilon_{\vec k n} / \partial k \vert_{k=k(\theta)}}  f(k(\theta),\theta).
\end{equation}
The angular integration is then replaced by a discrete sum as
\begin{equation}
	\int \frac{d^2 \vec k}{(2\pi)^2} \delta(E-\epsilon_{\vec k n}) f(\vec k) \rightarrow \sum_{\alpha, \theta_i} \frac{k_n^\alpha(\theta_i) \Delta \theta}{(2\pi)^2} \frac{1}{\vert \partial\epsilon_{\vec k n} / \partial k \vert_{k=k_{n}^{\alpha}(\theta_i)}} f(k_n^\alpha(\theta_i), \theta_i).
\end{equation}
The $k_i^\alpha(\theta)$ is the interpolated wave vector at the Fermi surface, which is characterized by the parameter $\theta$ (the corresponding angle of the momentum) for the $\alpha$ branch of the band $n$ Fermi surface.


Rewrite self-energies in \eqnref{eq:selfenergyfs} and vertex correction in \eqnref{eq:vertexcorrectionfs} with the interpolated Fermi surfaces as
\begin{equation}
	\Im \Sigma^R(k_n^\alpha(\theta_i), E) = n_i V_0^2 \sum_{n',\alpha', \theta_i'} \frac{k_{n'}^{\alpha'}(\theta_i') \Delta \theta}{(2\pi)^2} \frac{-\pi}{\vert \partial\epsilon_{\vec k n} / \partial k \vert_{k=k_{n'}^{\alpha'}(\theta_i')}} \lvert \bra{k_n^\alpha(\theta_i)} k_{n'}^{\alpha'}(\theta_i') \rangle \rvert^2.
	\label{eq:selfenergydiscrete}
\end{equation}
\begin{equation}
	\gamma_j(k_n^\alpha(\theta_i),E) = n_i V_0^2 \sum_{n',\alpha', \theta_i'} \frac{k_{n'}^{\alpha'}(\theta_i') \Delta \theta}{(2\pi)^2} \frac{-\pi}{\vert \partial\epsilon_{\vec k n} / \partial k \vert_{k=k_{n'}^{\alpha'}(\theta_i')}} \lvert \bra{k_n^\alpha(\theta_i)} k_{n'}^{\alpha'}(\theta_i') \rangle \rvert^2 \frac{1}{\Im \Sigma^R(k_{n'}^{\alpha'}(\theta_i'),E)} (J_j(k_{n'}^{\alpha'}(\theta_i')) +  \gamma_j(k_{n'}^{\alpha'}(\theta_i'),E)).
	\label{eq:vertexcorrectiondiscrete}
\end{equation}

\subsection{Current conservation}
\label{sec:currentconservation}

In this section, we focus on solving the vertex correction in \eqnref{eq:vertexcorrection} and \eqnref{eq:vertexcorrectionfs}.  We are going to show that the matrix to be inversed for vertex correction has zero determinant and its relation to current conservation.

The zero determinant of the matrix in  \eqnref{eq:gammatensor} to be inversed is illustrated with \eqnref{eq:selfenergydiscrete} and \eqnref{eq:vertexcorrectiondiscrete} of Fermi surface contributions.  By denoting
\begin{equation}
	\begin{split}
			M_{\vec k'n'} &= M_{k_{n'}^{\alpha'}(\theta_i')} = \frac{k_{n'}^{\alpha'}(\theta_i') \Delta \theta}{(2\pi)^2} \frac{-\pi}{\vert \partial\epsilon_{\vec k n} / \partial k \vert_{k=k_{n'}^{\alpha'}(\theta_i')}} \\
			V_{\vec k n, \vec k' n'} &= V(k_{n}^{\alpha}(\theta_i), k_{n'}^{\alpha'}(\theta_i')) = n_i V_0^2 \lvert \bra{k_{n}^{\alpha}(\theta_i)} k_{n'}^{\alpha'}(\theta_i') \rangle \rvert^2,
	\end{split}
\end{equation}
we have the self energies and the vertex correction as
\begin{gather}
	\Im\Sigma^R(\vec k n,E) = \sum_{\vec k' n'} V_{\vec k n, \vec k' n'} M_{\vec k'n'} \label{eq:selfenergydiscretesimple} \\
	\gamma_j(\vec k n, E)  = \sum_{\vec k' n'} V_{\vec k n, \vec k' n'} M_{\vec k'n'} \frac{1}{\Im\Sigma^R(\vec k' n',E)} (J_j(\vec k' n') + \gamma_j(\vec k' n',E)).
	\label{eq:vertexcorrectiondiscretesimple}
\end{gather}
The matrix  $V_{\vec k n, \vec k' n'}$ is symmetric.  The matrix to be inversed for the vertex correction is
\begin{equation}
	I_{\vec kn, \vec k'n'} = \delta_{\vec kn, \vec k' n'} - V_{\vec k n, \vec k' n'} M_{\vec k'n'} \frac{1}{\Im\Sigma^R(\vec k' n',E)}.
\end{equation}
For the singular value decomposition (SVD) of the matrix $I$, $I$ has a left eigen vector $e_{L0} = M_{\vec k n}$ and a right eigen vector $e_{R0} = \Im \Sigma(\vec k'n', E)$ as shown by
\begin{gather}
	\sum_{\vec k n} M_{\vec k n} I_{\vec kn, \vec k'n'} = M_{\vec k' n'} - (\sum_{\vec k n} M_{kn}  V_{\vec k n, \vec k' n'}) M_{\vec k'n'} \frac{1}{\Im\Sigma^R(\vec k' n',E)} = 0 \\
	\sum_{\vec k' n'}  I_{\vec kn, \vec k'n'}\Im \Sigma(\vec k'n', E) = \Im \Sigma(\vec kn, E) - \sum_{\vec k' n'}  V_{\vec k n, \vec k' n'} M_{\vec k'n'} = 0
\end{gather}
with \eqnref{eq:selfenergydiscretesimple}.
Both eigenvectors have the eigenvalue 0.  As a result, the matrix $I_{\vec kn, \vec k'n'}$ has zero determinant.

The existence of a solution to the vertex correction in \eqnref{eq:vertexcorrectiondiscretesimple} implies the current conservation physically. To see that, we re-write the vertex correction in the form of the SVD decomposition of the matrix $I$ as
\begin{equation}
\begin{pmatrix}
	e_{R0} & e_{R1}  & \cdots & e_{Rn}
\end{pmatrix}
\begin{pmatrix}
	0 &  &  & \\
	& \lambda_1 &  & \\
	&  & \ddots & \\
	&  &  & \lambda_n
\end{pmatrix}
\begin{pmatrix}
	e_{L0}^\dagger\\
	e_{L1}^\dagger\\
	\vdots\\
	e_{Ln}^\dagger
\end{pmatrix}
\gamma_j = \tilde{J_j}
\end{equation}
where $e_{Li} \ e_{Ri} \ \lambda_i$ are other left eigenvectors, right eigenvectors, eigenvales of $I$, statisfying $e_{Li}^\dagger e_{Rj} = \delta_{ij}$. The elements of $\gamma_j $ is $\gamma_j(\vec k n) $ in \eqnref{eq:vertexcorrectionfs} and the elements of $\tilde J_j$ is $ \sum_{\vec k' n'} V_{\vec k n, \vec k' n'} M_{\vec k'n'} J_j(\vec k' n') / \Im\Sigma^R(\vec k' n',E)$.  The condition of the existence of $\gamma_j$ is
\begin{equation}
	e_{L0}^\dagger \tilde{J_j} = \sum_{\vec k n ,\vec k' n'} M_{\vec k n} V_{\vec k n, \vec k' n'} M_{\vec k'n'} J_j(\vec k' n') / \Im\Sigma^R(\vec k' n',E) = \sum_{\vec k' n'} M_{\vec k'n'} J_j(\vec k' n') = 0.
\end{equation}
Restoring to the integral form, the condition becomes
\begin{equation}
	\int \frac{\dd^2 \vec k}{(2\pi)^2} \delta(E-\epsilon_{\vec k n}) J_j(\vec k n) = 0.
\end{equation}
Physically, this means the total current of the electrons in equilibrium is zero.  Since the charge $\rho$ is conserved, the current conservation $\partial_t \rho + \nabla \cdot \vec J=0$ is statisfied.

To solve the vertex correction, the inverse in \eqnref{eq:vertexcorrectiondiscretesimple} shall be replaced by Moore-Penrose pseudo-inverse \cite{barata2012moore}, given by
\begin{equation}
	\begin{pmatrix}
		e_{L0}^\dagger\gamma_j\\
		e_{L1}^\dagger\gamma_j\\
		\vdots\\
		e_{Ln}^\dagger\gamma_j
	\end{pmatrix}
	=
	\begin{pmatrix}
		0 &  &  & \\
		& 1/\lambda_1 &  & \\
		&  & \ddots & \\
		&  &  & 1/\lambda_n
	\end{pmatrix}
	\begin{pmatrix}
		e_{L0}^\dagger\tilde{J_j}\\
		e_{L1}^\dagger\tilde{J_j}\\
		\vdots\\
		e_{Ln}^\dagger\tilde{J_j}
	\end{pmatrix}
\end{equation}
in the SVD decomposition. Another way to solve for the vertex correction with this non-iterative method is using the finite frequency response $\chi_{ij}(\omega)$ and then decreasing $\omega$ for a convergent result in the $\omega \rightarrow 0$ limit.

\section{Numerical Results of Spin Transport}


\subsection{Rashba 2DEG}

\begin{figure*}[t]
	\centering
	\includegraphics[width=\textwidth]{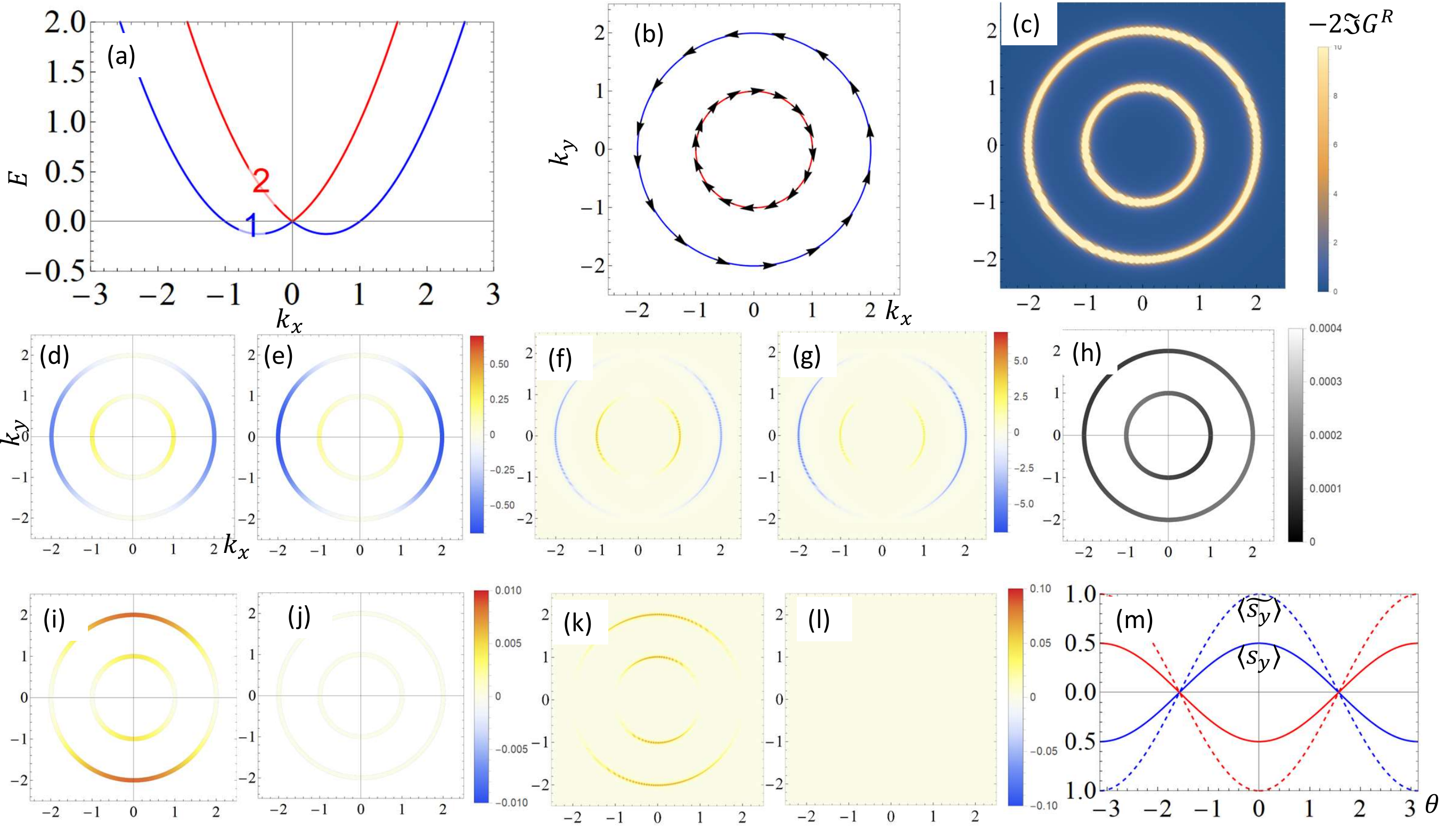}
	\caption{
		(a) The spectrum of Rashba 2DEG.
		(b) The Fermi surface for $E_f=1$ and the spin texture on the Fermi surface.
		(c) The spectral weight shown over the Brillouin zone.
		(d)(e) The contribution to CISP $\chi_{yx}$ on the Fermi surface of \eqnref{eq:cispfscontribution} without and with vertex correction.
		(e)(f) The contribution to CISP $\chi_{yx}$ over the Brillouin zone of \eqnref{eq:cispcontribution} without and with vertex correction.
		(h) the relaxation time over the Fermi surfaces $\tau_{\vec k n}+20$.
		(i)(j) The contribution to SHE $\sigma^z_{yx}$ on the Fermi surface of \eqnref{eq:shefscontribution} without and with vertex correction.
		(k)(l) The contribution to SHE $\sigma^z_{yx}$ over the Brillouin zone  of \eqnref{eq:shecontribution} without and with vertex correction.
		(m) The y component of the spin directions on the Fermi surface.  The solid (dashed) lines are the spin vertex without (with) the vertex correction.  The spin components are shown in the polar angle $\theta$ of the Brillouin zone.
	}
	\label{fig:Rashba2DEG}
\end{figure*}

To validate our numerical methods, we will first evaluate the spin transport of 2-dimensional electron gas (2DEG) with Rashba spin-orbit coupling, which can be directly compared with the analytical solutions in Ref. [\onlinecite{inoue2003diffuse}].

The Hamiltonian of the Rashba 2DEG is
\begin{equation}
	H(k) = k^2 / 2 + \alpha_R (k_y \sigma_x - k_x \sigma_y).
\end{equation}
\figref{fig:Rashba2DEG}(a) shows the spectrum of two spin splitting bands, with the parameters $\alpha_R=0.5$ and $E_f=1$. \figref{fig:Rashba2DEG}(b) shows the Fermi surfaces and the spin texture on the Fermi surfaces.  The Fermi surfaces are two concentric circles.  The spins on the Fermi surfaces of band 1 and 2 are opposite. \figref{fig:Rashba2DEG}(c) shows  The spectral weight $-2\Im Tr G^R(\vec k,E_f)$ with $n_iV_0^2=0.05$.  The two bands are clearly resolved in the dilute disorder limit.

We directly compare our numerical results for CISP with theoretical values obtained from Ref. [\onlinecite{inoue2003diffuse}] for a typical set of parameters listed above. The CISP from the full numerical calculations of the formula \eqnref{eq:cisp} is $-0.79878$ without vertex correction and $-1.5865$ with vertex correction. The simplified formula \eqnref{eq:cispfs} only with the Fermi surface contribution gives $-0.79572$ without vertex corrections and $-1.5914$ with vertex correction. Thus, the two results from different expressions give nearly identical the CISP response coefficients that also agree with the theoretical values $-\alpha_R/ 4 \pi n_iV_0^2 = -0.79578$ without and $-\alpha_R/ 2 \pi n_iV_0^2=-1.5916$ with vertex corrections from Ref. [\onlinecite{inoue2003diffuse}] within $1\%$
.  The error can be reduced by taking denser meshes.

Next we provide an intuitive physical picture for the general understanding of the origin of the CISP $\chi_{yx}$. Under the electrical field along the $+x$ direction, the electrons are in the non-equilibrium states and the corresponding Fermi sphere shifts to the $-x$ direction.  For the Fermi surfaces of the band 1 in \figref{fig:Rashba2DEG} (b)
, less electrons occupy the states with $k_x>0$ and spin directions $s_y>0$ while more electrons occupy the states with $k_x<0$ and spin directions $s_y<0$.  As a result, a charge current in $+x$ direction gives rise to a negative contribution to the CISP $\chi_{yx}$ ($s_y<0$) for the Fermi surface of the band 1, as shown by the blue color in \figref{fig:Rashba2DEG}(d)(e)(f)(g)
.  Thus, the electrons on this Fermi surface contributes $\propto (k_f + 2 \alpha_R) \tau$. $\tau$ is the relaxation time and constant over the whole Fermi surfaces for both bands as shown in \figref{fig:Rashba2DEG}(h) with values close to the theoretical one $-n_iV_0^2/2$.  The electrons on the Fermi surface of the band 2 gives a positive contribution (red color) to CISP due to the opposite spin texture as shown in \figref{fig:Rashba2DEG}(d)(e)(f)(g), which is proportional to $k_f \tau$. Adding the contributions from two Fermi surfaces, we find $\chi_{yx} \propto \alpha_R\tau$.

For the vertex correction, it is more convenient to consider the correction to spin operator instead of current operator. Theoretically, the spin operator with the vertex correction $\bra{\vec k n} \tilde{S}_y(E) \ket{\vec k n}$ in \eqnref{eq:spinvertexcorrectionfs} doubles the bare spin operator matrix elements $\bra{\vec k n} S_y \ket{\vec k n}$. Thus, CISP with the vertex correction is $\tilde{\chi}_{yx} = 2 \chi_{yx}$.  \figref{fig:Rashba2DEG}(m) shows the numerical spin operator matrix element from \eqnref{eq:spinvertexcorrectionfs}
.  The spin operator with the vertex correction doubles the bare spin operator over the whole Fermi surfaces, consistent with the theoretical results.

Finally, we show that the numerical results for SHE also agree with theoretical results. The calculated SHE $\sigma^z_{yx}$ from the formula \eqnref{eq:she} are $0.0397802$ without vertex correction and $1.26112 \times 10^{-4}$ with vertex correction.  The simplified formula \eqnref{eq:shefs} gives $0.039789$ without vertex correction and $-2.75749\times 10^{-4}$ with vertex corrections.  Both agree with each other and are consistent with the theoretical values $1/8\pi$ without vertex correction and $0$ with vertex correction from \refcite{inoue2004suppression} within the error $1\%$.

\figref{fig:Rashba2DEG}(i)-(l) shows the contribution to the response coefficient $\sigma^z_{yx}$ of SHE defined by \eqnref{eq:shefscontribution} and \eqnref{eq:shecontribution}.  Without vertex correction, the off-diagonal terms contribute a positive value to SHE (red color), which is maximal on the y axis and nearly zero on the x-axis.  The diagonal contribution is almost zero as $\bra{\vec k n} J_y^z \ket{\vec k n}$ is zero.  When including the vertex correction, we find the SHE coefficient $\sigma^z_{yx}$ is almost zero, as predicted by the analytical theory
This reveals the cancellation of the intrinsic and intrinsic-side-jump contributions to SHE \cite{sinova2015spin}.


\subsection{Spin texture of monolayer Pb}

\begin{figure*}[t]
	\centering
	\includegraphics[width=\textwidth]{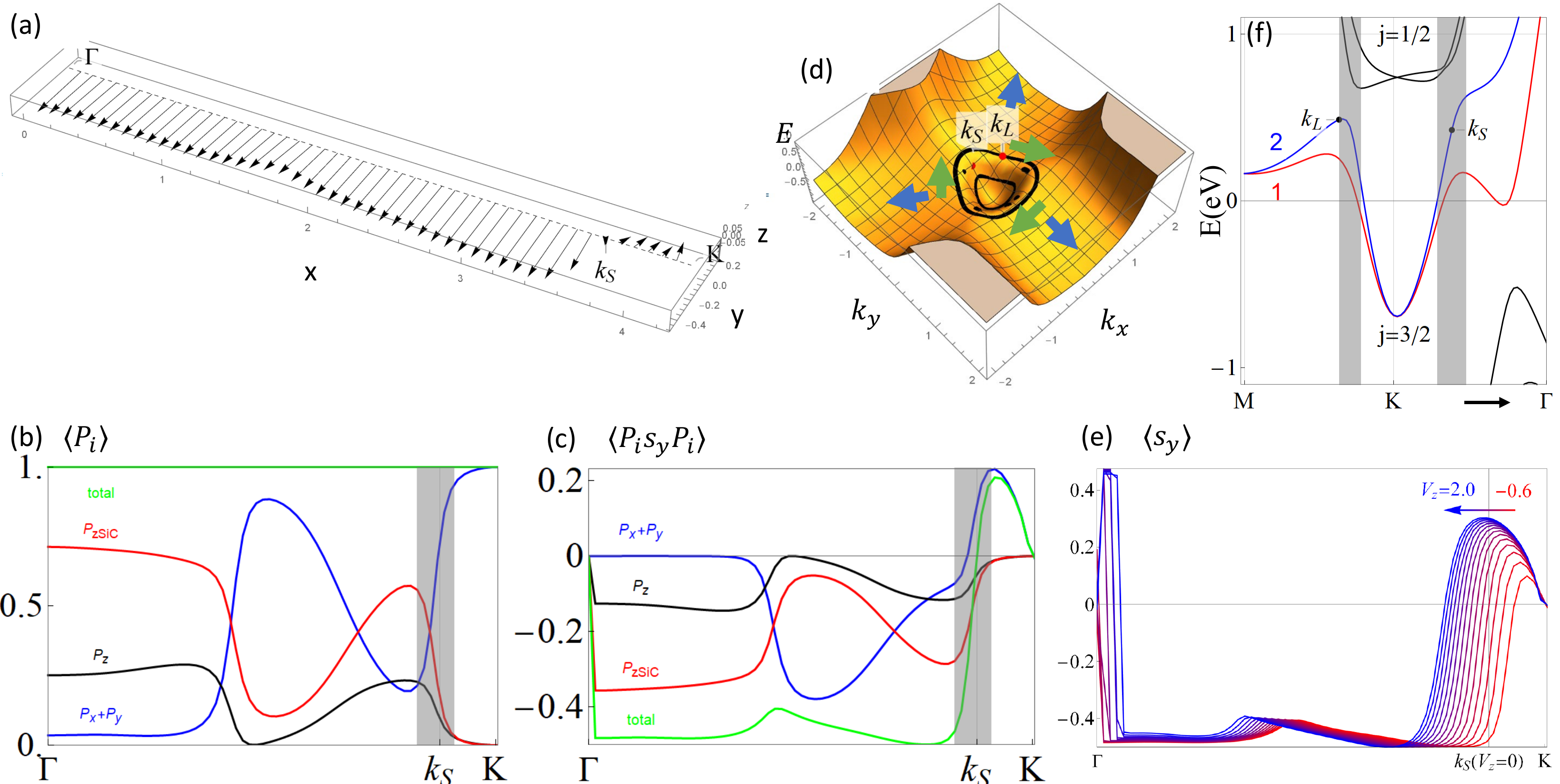}
	\caption{
		(a) Spin directions along the $\Gamma-K$ line in 3D.
		(b) The projection of states along  $\Gamma-K$  to different orbitals.  The total is the sum of all orbitals.  The grey region is the anti crossing region.
		(c) The contribution to y components of spin directions from different orbitals.
		(d) The energy spectrum over 2D Brillouin zone around $K$ in 3D.
		%
		(e) The y components of spin directions from all orbitals for different $V_z$ from $-0.6 eV$ to $2.0 eV$ with an interval $0.2eV$ indicated by the colorful arrow. 
		(f) The energy spectrum with $V_z =1 eV$.
	}
	\label{fig:MSAV}
\end{figure*}

In this section, we will discuss the physical properties and origins of the momentum-space spin anti-vortex, as well as the relation to the Lifshitz transition of Fermi surfaces.

First, we show the matrix element $\bra{\vec k n} S_i \ket{\vec k n}$ of spin operator at the location of spin anti-vortex approaches zero.  \figref{fig:MSAV}(a) shows the spin direction along the $\Gamma-K$ line.  At the high symmetry points $\Gamma$ and $K$, the states are degenerate and the spin directions are not well defined.
We find the spin directions along this line mainly lies in the plane and point to the $y$ axis as a consequence of the $x$-directional mirror
and time reversal symmetries. As the spin directions near $\Gamma$ and $K$ are opposite, it has to approach zero at certain momentum in the $\Gamma$-$K$ line, which is labelled as $k_S$ in \figref{fig:MSAV}(a).


The spin anti-vortex originates from the anti-crossing between the bands with opposite spin directions. \figref{fig:MSAV}(b) shows the projection of eigen-states to different orbitals, defined as $\bra{\vec k n} P_i \ket{\vec k n}$ with the projection operator $P_i = \ket{p_i} \bra{p_i}$ to the orbital $p_i$, along $\Gamma-K$. Around $k_S$, the dominant orbital nature of the bands changes from the $p_z$ orbital of SiC (red line) to the $p_x$ and $p_y$ orbitals of Pb (blue lines). As the orbital nature varies rapidly across the anti-crossing regime, we choose the following condition (the gradient of the probability difference between different orbitals)
\begin{equation}
	\lvert \nabla_\vec k \bra{\vec k n} P_x+P_y - P_{zSiC} \ket{\vec k n} \rvert > 1
\end{equation}
to determine the anti-crossing regime, as shown by the grey region in \figref{fig:MSAV}(b) and (c).  \figref{fig:MSAV}(c) shows the contribution to $s_y$ components from different orbitals $\bra{\vec k n} P_i s_y P_i \ket{\vec k n}$.  The $p_{zSiC}$ orbital carries negative $s_y$ while the $p_x$ and $p_y$ orbitals carry positive $s_y$.  As the orbital nature of the bands varies from the $p_{zSiC}$ orbital to the $p_x$ and $p_y$ orbitals, zero $s_y$ occurs in the anti-crossing regime and determines the location of the spin anti-vortex $k_S$.

Besides spin anti-vortex, the band anti-crossing also leads to the Lifshitz transition of Fermi surfaces.  The Lifshitz transition happens at the saddle points of the spectrum.  \figref{fig:MSAV}(d) shows the spectrum around $K$ in the 2D BZ and the saddle point is labeled by $k_L$.
In the anti-crossing region between two black lines, the spectrum shows a loop of local maxima in the radial direction. Along the tangential direction of this loop, the red points are the local minima and these minima define the saddle points in the energy spectrum. The Lifshitz transition happens at these saddle points. In summary, both spin anti-vortex and Lifshitz transition happen in the anti-crossing region as shown in \figref{fig:MSAV}(d).

 In addition, the location of the spin anti-vortex can controlled by an external gate that tunes the difference in potentials between the Pb layer and SiC substrate. To model the effect, we include a constant potential $V_z$ to the $p_z$ orbital of SiC and the energy spectrum for $V_z=1.0$eV is shown in \figref{fig:MSAV}(f).  Compared to the Fig. 2(d) in the main text for $V_z=0$eV, the $j=1/2$ bands from the SiC for $V_z=1$eV have higher energies and the band anti-crossing region moves away from $K$ point, which leads to the change of the location of both the Lifshitz transition point $k_L$ and the spin anti-vortex point $k_S$.  To show the movable range of the spin anti-vortex, we plot the y component of spin directions on the $\Gamma-K$ line from all orbitals for different $V_z$ in \ref{fig:MSAV}(e). The zeros of these lines around $k_S(V_z=0)$ indicate the locations of spin anti-vortex points for different $V_z$.  The spin anti-vortex could be moved towards $\Gamma$ by positive $V_z$ (blue lines) or $K$ by negative $V_z$ (red lines).

\subsection{Numerical CISP for monolayer Pb}

\begin{figure*}[t]
	\centering
	\includegraphics[width=\textwidth]{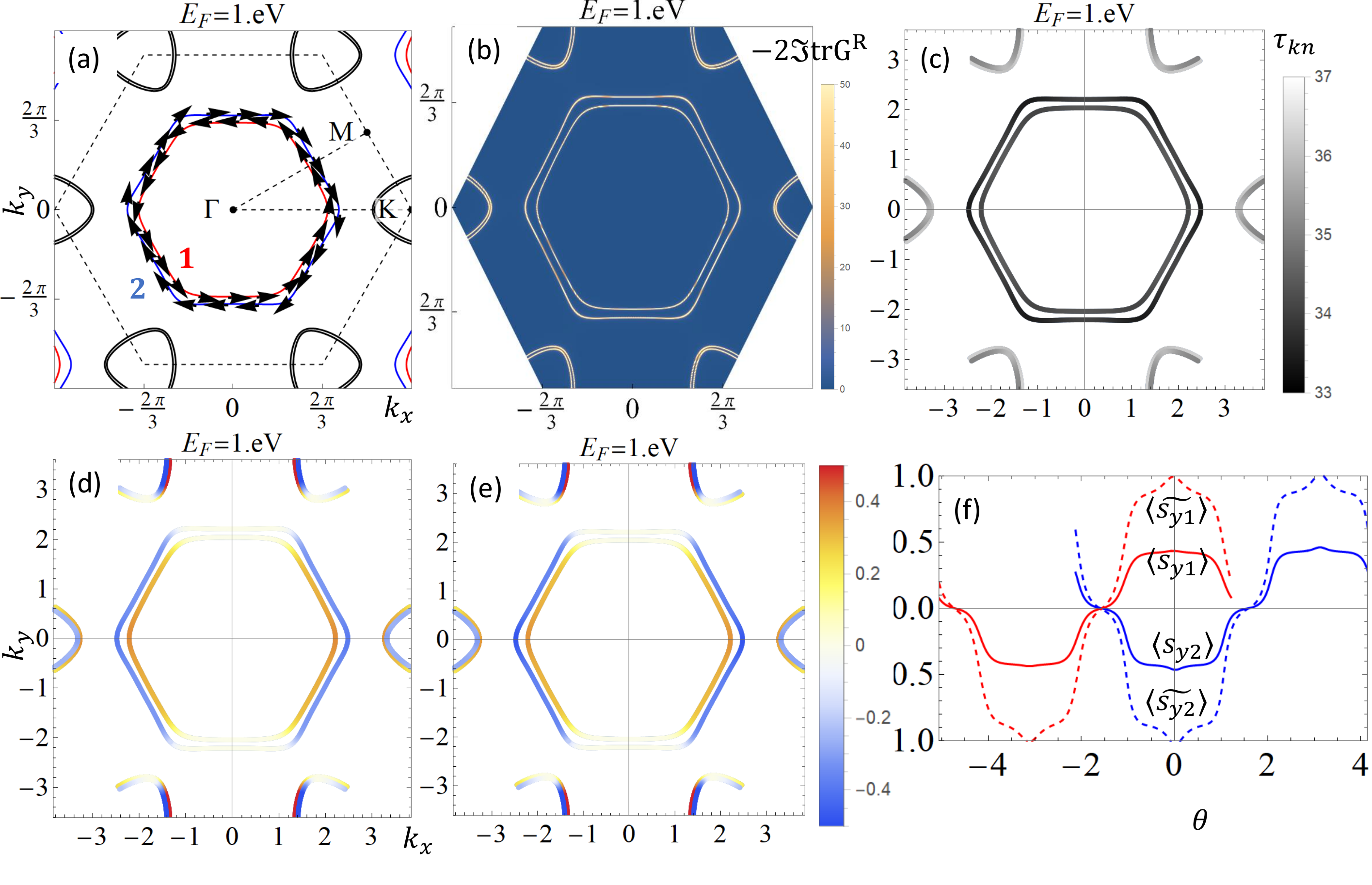}
	\caption{
		(a) Fermi surfaces and spin texture for $E_F=1eV$.
		(b) Spectral weight for $E_F=1eV$ over the Brillouin zone.
		(c) Relaxation time over the Fermi surfaces.
		(d)(e) The contribution to CISP on Fermi surfaces without and with vertex correction.
		(f) The $s_y$ over Fermi surfaces for band 1 and 2.  The red line (blue) is band 1 (2).  The solid line is the bare spin vertex. The dashed line is the corrected spin vertex.
	}
	\label{fig:CISPhighEF}
\end{figure*}

\begin{figure*}[t]
	\centering
	\includegraphics[width=\textwidth]{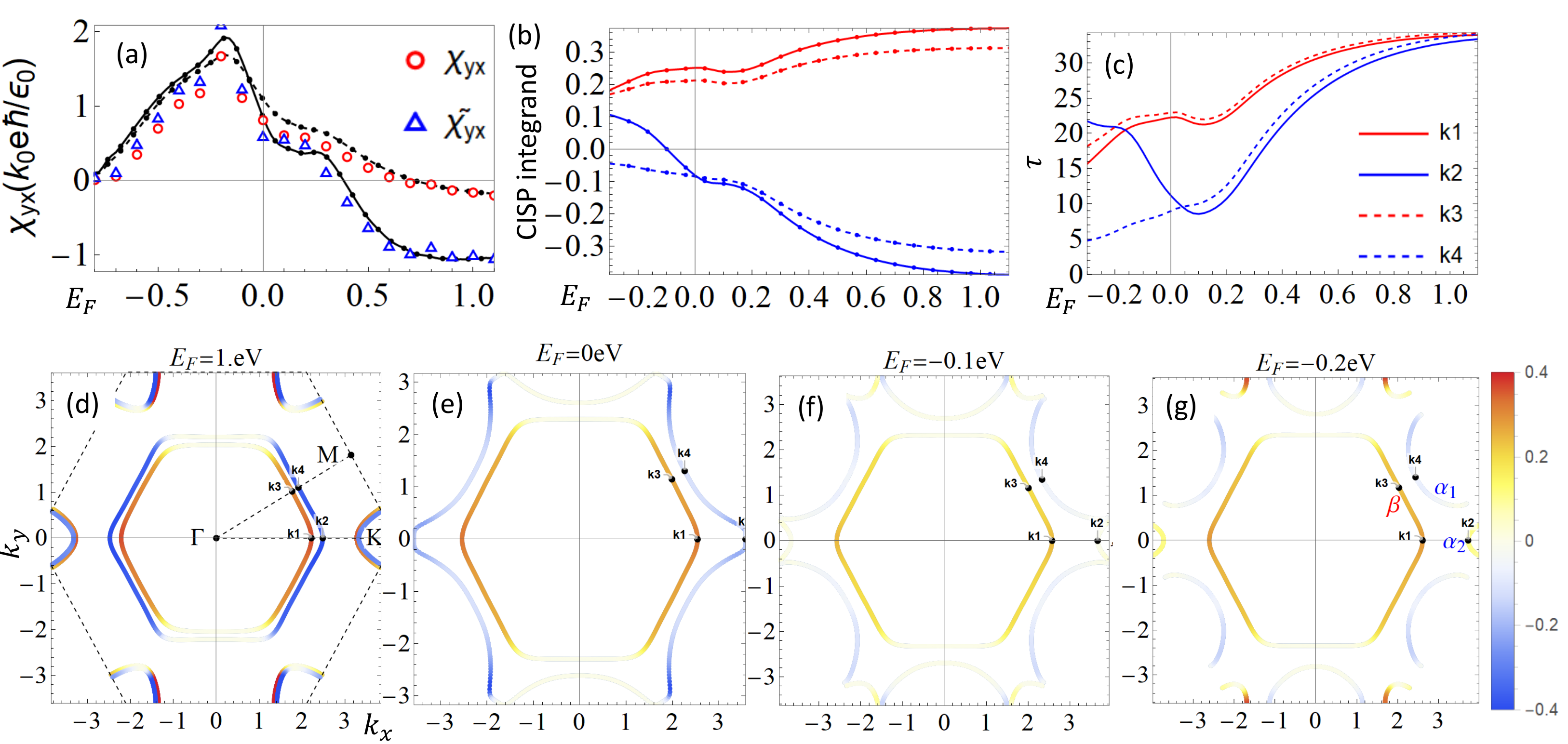}
	\caption{
		(a) CISP.  The solid (dashed) line is calculated from the Brillouin zone formula \eqnref{eq:cisp} with (without) vertex correction.  The red circles (blue triangles) are calculated from the Fermi surface formula \eqnref{eq:cispfs}.
		(b)(c) The contribution to CISP and relaxation time at the intersection points $k_1 k_2 k_3 k_4$ between the Fermi surfaces of band 1 and 2 and the high symmetry lines $\Gamma-K$ and $\Gamma-M$ lines.
		(d)(e)(f)(g) the contribution to CISP over the Fermi surfaces \eqnref{eq:cispfscontribution}  for $E_F=1eV,0eV,-0.1eV,-0.2eV$.
	}
	\label{fig:CISPEF}
\end{figure*}

In this section, we will focus on understanding the numerical results of CISP for monolayer Pb by comparing with Rashba 2DEG at high $E_f$ and explaining the behaviors of CISP when varying $E_f$. The numerical calculation by \eqnref{eq:cispfs} at $E_F=1eV$ for the band 1 and 2 in \figref{fig:CISPhighEF}(a)
is consistent with the physical picture of Fermi surfaces shift in Rashba 2DEG. To see that,  \figref{fig:CISPhighEF}(a) shows the spin texture on the Fermi surfaces of bands 1 and 2, similar to Rashba 2DEG and opposite to each other.
All the calculations of CISP are done within dilute disorder limits, for which the band broadening induced by disorder is smaller than spin splitting energies and the spin splitting bands are well resolved in the spectral weight as shown in \figref{fig:CISPhighEF}(b). The corresponding relaxation time is shown in \figref{fig:CISPhighEF}(c), and we find the relaxation time to be uniform for each Fermi surface, but its values for the bands 1 and 2 are slightly different. 
Another difference from Rashba 2DEG is the Fermi surfaces of band 1 and 2 are hole pockets, not the electron pockets.  When electrical fields are applied in the $+x$ direction, rather than less electrons with $k_x>0$ and $s_y<0$ for electron pockets,  there are more electron with $k_x>0$ and $s_y<0$ and less electrons with $k_x<0$ and $s_y>0$.  The contribution from this Fermi pocket to CISP is negative (blue color) as shown in \figref{fig:CISPhighEF}(d)(e).  The red Fermi surface has the opposite spin texture and contribute positively (red color) as shown in \figref{fig:CISPhighEF}(d)(e).
As Rashba 2DEG, the contribution from the blue Fermi pockets is canceled partly by the red one and make a total negative contribution as shown in \figref{fig:CISPEF}(a) at $E_f=1eV$.

Under $\TR$ symmetry, the contribution to CISP in \eqnref{eq:cispfscontribution} becomes
\begin{equation}
	\begin{split}
		\chi_{ij} (\vec k n) & =  -\frac{1}{ 2\pi } \langle \vec k n \vert S_i \vert \vec k n \rangle \langle \vec k n \vert \Gamma_j(\vec k, E_f) \vert \vec k n \rangle \delta(E_f -\epsilon_{\vec k n}) \frac{\pi}{-\Im \Sigma^R(\vec k n, E_f)} \\
		& =  -\frac{1}{ 2\pi } \langle - \vec k n \vert(\TR S_i \TR^\dagger)\vert - \vec k n \rangle^* \langle - \vec k n \vert (\TR \Gamma_j(\vec k, E_f) \TR^\dagger ) \vert - \vec k n \rangle^* \delta(E_f -\epsilon_{\vec k n}) \frac{\pi}{-\Im \Sigma^R(\vec k n, E_f)} \\
		& =   -\frac{1}{ 2\pi }  \langle - \vec k n \vert(- S_i )\vert - \vec k n \rangle^* \langle - \vec k n \vert ( -\Gamma_j(\vec k, E_f) ) \vert - \vec k n \rangle^* \delta(E_f -\epsilon_{-\vec k n}) \frac{\pi}{-\Im \Sigma^R(-\vec k n, E_f)} \\
		& = \chi_{ij} (-\vec k n).
	\end{split}
	\label{eq:CISPTR}
\end{equation} 
This means the contribution to CISP should be the same at $\vec k$ and $-\vec k$, which is clearly shown by the numerical calculations in \figref{fig:CISPhighEF}(d)(e) and \figref{fig:CISPEF}(d)-(g).

We also find that the vertex correction increases the matrix elements of the spin operator, similar to the case of Rashba 2DEG.  \figref{fig:CISPhighEF}(f) shows the corrected spin operator in \eqnref{eq:spinvertexcorrectionfs} is proportional to and larger than the bare spin operator and share similar profiles over $\theta$.  

The numerical calculations of the CISP from \eqnref{eq:cispfs} and \eqnref{eq:cisp} are shown by solid (dashed) lines and circles (triangles) in \figref{fig:CISPEF}(a) for the cases with (without) vertex corrections for different $E_F$. Interestingly, we notice a sign change of the CISP when lowering $E_F$ from $1eV$ to $0eV$ and a rapid change around $E_F=-0.1eV$.
The sign change originates from the different relaxation time for the bands 1 and 2.  \figref{fig:CISPEF}(c) shows the relaxation time on four typical points $k_1$ to $k_4$ for different $E_F$.  The four points are intersection points between Fermi surfaces of band 1 and 2 with the high symmetry lines $\Gamma-K$ and $\Gamma-M$.  \figref{fig:CISPEF}(b) shows the contribution to CISP at these points.  Around $E_F=1eV$, the relaxation time for all the momenta in \figref{fig:CISPEF}(c) has a similar magnitude, which indicates a similar relaxation time for the bands 1 and 2. Correspondingly, the sign of the CISP is determined by the Fermi surface size of the bands 1 and 2, and since the band 2 has a large Fermi surface size, the CISP coefficient $\chi_{yx}$ has a negative sign.
When lowering $E_F$ to $0eV$, the relaxation time of the band 2 (blue lines) decreases more rapidly than that of the band 1 (red lines) in \figref{fig:CISPEF}(c), which leads to a dominant contribution from the band 1 for the CISP and thus, $\chi_{yx}$ changes its sign to be positive around $E_F=0eV$.


Besides the sign change, we also notice a rapid change in CISP coefficient $\chi_{yx}$ between $E_F=0eV$ and $E_F=-0.2eV$, which is closely related to the spin anti-vortex around $E_F=-0.1eV$.  When lowering $E_F=0eV$ to $E_F=-0.2eV$, the Fermi pocket $\alpha_2$ around the $K$ point comes across spin anti-vortex at $k_S$.  The spin texture at two sides of the spin anti-vortex are opposite along the $\Gamma-K$ line as shown in \figref{fig:MSAV}(a). Furthermore, the $\alpha_2$ Fermi pockets in Fig. \figref{fig:CISPEF}(g) have a relatively large relaxation time. This means the contribution of CISP from the Fermi pockets $\alpha_2$ of the band 2 has the same sign as that from the band 1, thus leading to a rapid increasing of the CISP around $E_F=-0.1eV$ in \figref{fig:CISPEF}(a).

The rapid change in CISP shows a special feature of spin anti-vortex that it is located at non-high-symmetry momenta when compared to the spin vortex texture located at time reversal invariant momenta in the Rashba model.  This is explained as follows. First, we can consider a time reversal invariant momentum. Such a momentum is always enclosed by a Fermi pocket since opposite momenta around a time reversal invariant momentum have to be both occupied or unoccupied; e.g. $\Gamma$ is always enclosed by the Fermi surface $\beta$ independent of $E_\text F$, unless the whole Fermi surface completely disappears. Next, we consider a spin anti-vortex that is located at a non-high-symmetry momentum. Consequently, it allows the Fermi surface to pass through the anti-vortex core as $E_\text F$ varies: e.g. as $E_\text F$ lowers in \figref{fig:CISPEF}(e), the Fermi surface $\alpha$ (red) passes through $k_S$ and evolves into the Fermi surface $\alpha_2$ (green). The Fermi surfaces before and after passing through the spin anti-vortex core have opposite spin textures and thus the contribution from the spin anti-vortex to the CISP  can change its sign rapidly. \figref{fig:CISPEF}(e) and (g), where $E_\text F$ is lowered and the CISP contribution from the Fermi surface near $k_S$ changes from negative (blue, from the Fermi surface $\alpha$) to positive (red, from the Fermi surface $\alpha_2$). In comparison, the spin polarization at the Fermi pocket $\beta$ around $\Gamma$ in \figref{fig:CISPEF}(e) and (g) keeps the positive sign (red) and can only vary relatively slowly.


\subsection{Numerical SHE for monolayer Pb}

\begin{figure*}[t]
	\centering
	\includegraphics[width=\textwidth]{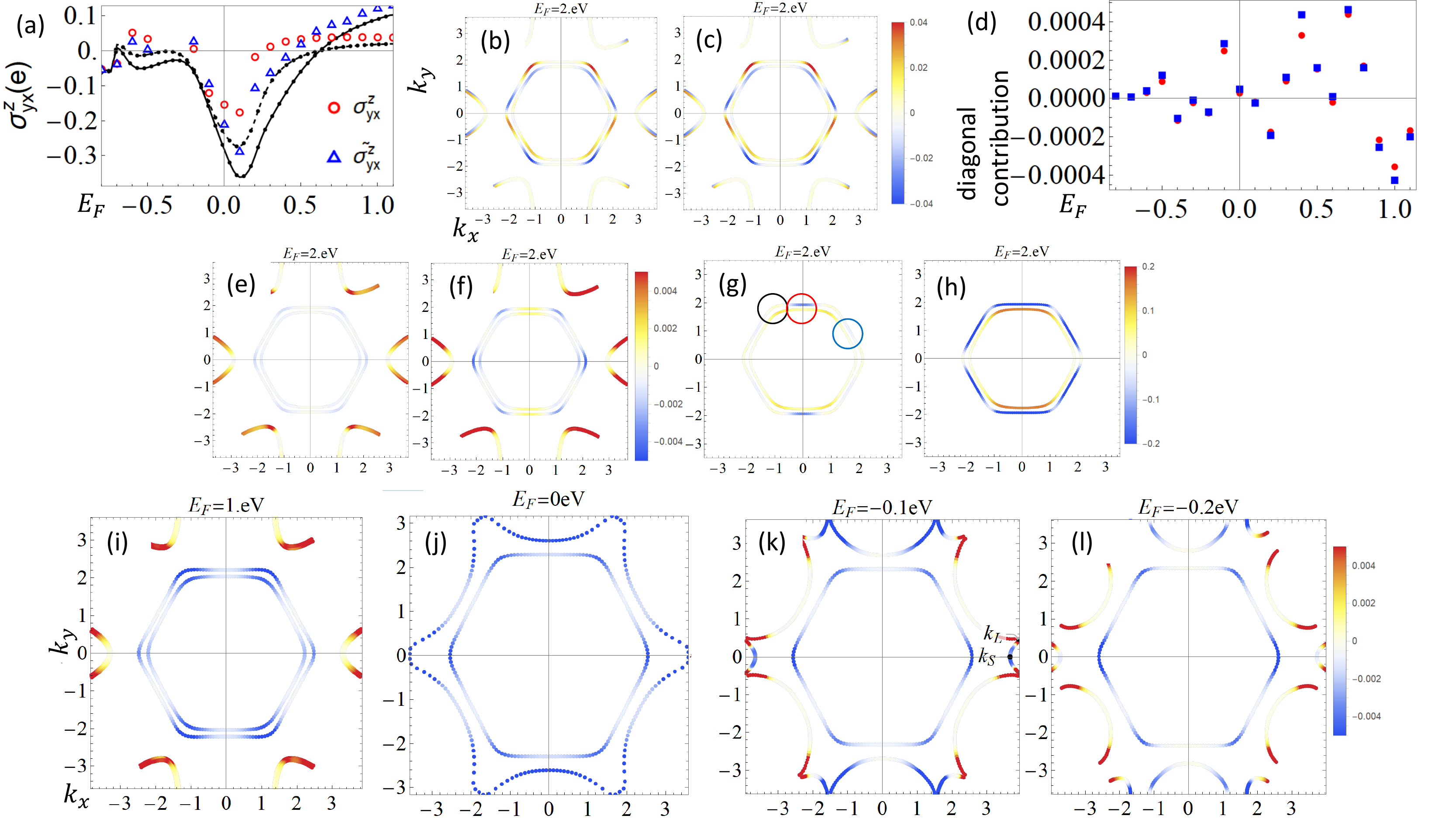}
	\caption{
		(a) SHE.  The solid (dashed) line is calculated from the Brillouin zone formula \eqnref{eq:she} with (without) vertex correction.  The red circles (blue triangles) are calculated from the Fermi surface formula \eqnref{eq:shefs}.
		(b)(c) The diagonal contribution to SHE over Fermi surfaces at $E_F=2eV$ without and with vertex corrections.
		(d) The diagonal contribution to SHE for different $E_F$.
		(e)(f) The off diagonal contribution to SHE over Fermi surfaces at $E_F=2eV$ without and with vertex corrections.
		(g)  $\Im (\langle \vec k n' \vert J^k_i \vert \vec k n \rangle \langle \vec k n \vert J_j(\vec k,E_f) \vert \vec k n' \rangle)$ over Fermi surfaces of $E_F=2eV$.
		(h)  $\Im (\langle \vec k n' \vert J^k_i \vert \vec k n \rangle \langle \vec k n \vert \Gamma_j(\vec k,E_f) \vert \vec k n' \rangle)$ including vertex corrections.
		%
		%
		(i)(j)(k)(l) the off diagonal contribution to SHE over the Fermi surfaces for $E_F=1eV, 0eV,-0.1eV,-0.2eV$ with vertex corrections.
	}
	\label{fig:SHE}
\end{figure*}

In this section, we discuss the numerical calculations of the SHE coefficient of monolayer Pb shown in \figref{fig:SHE}(a).  For $\sigma^z_{yx}$ at high $E_F$, we check our calculation and compare with Rashba 2DEG.  When lowering $E_F$ from $1eV$ to $0eV$, there is a sign change of $\sigma^z_{yx}$ and for $E_F$ between $0.1eV$ and $-0.2eV$,  $\sigma^z_{yx}$ are changing rapidly, which are explained next.

To validate our results, we first examine $\TR$ symmetry property of the numerical results. \figref{fig:SHE}(b) and (c) show the diagonal contribution to the SHE (\eqnref{eq:shetimereversal}) at Fermi surfaces, which is opposite at $\vec k$ and $-\vec k$, as required by $\TR$.
When summed over Fermi surfaces, the diagonal contribution shown in \figref{fig:SHE}(d) is negligible as \eqnref{eq:shetimereversal} expected.
It can be made closer to zero by taking more points on Fermi surfaces.
For off-diagonal contribution ($n'\neq n$ in the equation below), $\TR$ symmetry gives
\begin{equation}
\begin{split}
	\sigma^k_{ij}(\vec kn) &= -\frac{1}{2\pi} \sum_{n'\neq n} 2 \pi \delta(E_f - \epsilon_{\vec k n}) \frac{1}{\epsilon_{\vec k n} - \epsilon_{\vec k n'}} \Im (\langle -\vec k n' \vert (\TR J^k_i \TR^\dagger )\vert -\vec k n \rangle^* \langle - \vec k n \vert (\TR \Gamma_j(\vec k,E_f) \TR^\dagger) \vert -\vec k n' \rangle^*). \\
	& =  -\frac{1}{2\pi} \sum_{n'\neq n} 2 \pi \delta(E_f - \epsilon_{\vec k n}) \frac{1}{\epsilon_{\vec k n} - \epsilon_{\vec k n'}} (-)\Im (\langle -\vec k n' \vert ( J^k_i )\vert -\vec k n \rangle \langle - \vec k n \vert (- \Gamma_j(\vec k,E_f) ) \vert -\vec k n' \rangle)\\
	& =  -\frac{1}{2\pi} \sum_{n'\neq n}  2 \pi \delta(E_f - \epsilon_{-\vec k n}) \frac{1}{\epsilon_{-\vec k n} - \epsilon_{-\vec k n'}} (-)\Im (\langle -\vec k n' \vert ( J^k_i )\vert -\vec k n \rangle \langle - \vec k n \vert (- \Gamma_j(\vec k,E_f) ) \vert -\vec k n' \rangle) \\
	& = \sigma^k_{ij}(-\vec kn).
\end{split}
\end{equation}
So the off diagonal contribution should share the same color at opposite momenta, consistent with the calculated one shown in \figref{fig:SHE}(e)(f).



The cancellationg between intrinsic and intrinsic-side-jump contribution in Rashba 2DEG \cite{sinova2015spin} does not happen for monolayer Pb because the band 1 and 2 at large $E_F$ is quantatively different from the Rashba model.
\figref{fig:SHE}(g) and (h) show $\Im (\langle \vec k n' \vert J^k_i \vert \vec k n \rangle \langle \vec k n \vert \Gamma_j(\vec k,E_f) \vert \vec k n' \rangle)$ in \eqnref{eq:shefs} without and with vertex correction. In contrast, for Rashba 2DEG, it is known that
\begin{equation}
	\Im (\langle \vec k n' \vert J^k_i \vert \vec k n \rangle \langle \vec k n \vert J_j(\vec k,E_f) \vert \vec k n' \rangle) = \frac{v k_y^2}{2 k}
\end{equation}
without vertex corrections and $0$ including vertex corrections. Thus, it is clear that this quantity $\Im (\langle \vec k n' \vert J^k_i \vert \vec k n \rangle \langle \vec k n \vert \Gamma_j(\vec k,E_f) \vert \vec k n' \rangle)$ shows different behaviors for monolayer Pb and Rashba model.



We next consider the sign change in \figref{fig:SHE}(a).  It can be understood from the disappearance of the two electron Fermi pockets with positive contribution (red color) in \figref{fig:SHE}(i).
At $E_F=1eV$, the electron Fermi pockets dominate, giving positive contribution.  At $E_F=0eV$, the electron Fermi pockets disappear and the hole Fermi pocktets of band 1 and 2 gives negative contribution.  So when lowering $E_F$ from $1eV$ to $0eV$, there is a sign change of $\sigma_{yx}^z$.

The rapid change is related to the Lifshitz transition of Fermi surfaces of band 2 happened at $k_L$ around $E_F=-0.1eV$.  At $E_F=-0.1eV$, the contribution on Fermi surfaces around $k_L$ switches from negative (blue) to positive (red) values as shown in  \figref{fig:SHE}(j)(k)(l).  The contribution remains negative around spin anti-vortex.
Because contribution around $k_L$ change its sign in a narrow energy range, the total contribution has a large variation.

\subsection{Estimate of the experimental observables}

\begin{figure*}[t]
	\centering
	\includegraphics[width=\textwidth]{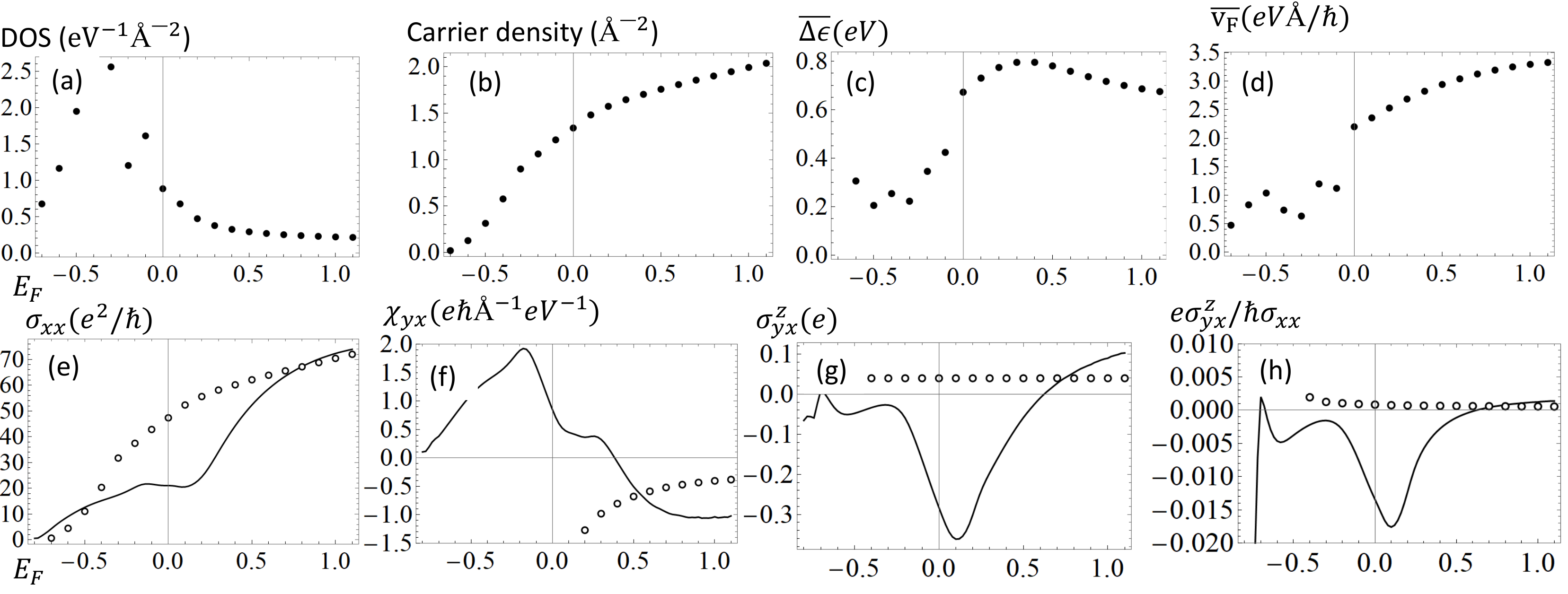}
	\caption{
		(a)(b)(c)(d) Density of states, carrier density, average of spin splitting energy, average of Fermi velocity with band 1 and 2 for different $E_F$
		(e)(f)(g)(h) comparison between numerical results of monolayer Pb and Rashba 2DEG in longitudinal conductivity, CISP, SHE, spin hall angle over $E_F$.  The solid line is the numerical response functions with vertex corrections. The black circles are calculated with the Rashba 2DEG model.
	}
	\label{fig:compareRashba}
\end{figure*}

In this section, we will give a numerical estimate of the magnitude of $\chi_{yx}$ and $\sigma^z_{yx}$ and compare the values with those estimated from Rashba 2DEG.
In the estimate below, we choose the parameter combination $n_i V_0^2 = 0.05 eV^2\AA^{-2}$ to characterize the impurity density and strength.

To compare with Rashba 2DEG, the mobility $\mu$ is extracted from the longitudinal conductivity $\sigma_{xx} = n e \mu
$.  $n = \sum_{n \vec k} \theta(E_F-\epsilon_{\vec k n})$ is the carrier density.
The effective mass is evaluated from the density of states $D = m / \pi \hbar^2$ and the Rashba SOC strength is evaluated from spin splitting energy $\Delta\epsilon / 2 k_F$ averaged over Fermi surfaces of band 1 and 2.  $\Delta \epsilon = \epsilon_{\vec k_F, 4} - \epsilon_{\vec k_F, 3}$ and $k_F$ is combined with $m$ for the averaged Fermi velocity $v_F =  k_F / m$.

The comparisons between monolayer Pb and Rashbs 2DEG are shown in \figref{fig:compareRashba} and they share same order of magnitude.  The mobility is fitted to be $\sim 50 cm^2 /Vs$ from $\sigma_{xx}$.  Other response quantities for Rashba 2DEG are calculated in the following way
\begin{equation}
	\chi_{yx} = -\frac{1}{2\pi} \tau \alpha_R \frac{me}{\hbar} = -\frac{1}{4 } D \mu \frac{\Delta \epsilon}{v_F} 
\end{equation}
\begin{equation}
	\sigma^z_{yx} = \frac{e}{8\pi}
\end{equation}
The spin hall angle is
\begin{equation}
	\frac{e}{\hbar} \frac{\sigma^z_{yx}}{\sigma_{xx}} = \frac{e}{\hbar} \frac{1}{8\pi n \mu}
\end{equation}

%